\documentclass[preprint,12pt]{elsarticle}

\usepackage{graphics}
\usepackage{graphicx}

\journal{Physica A}

\begin{document}

\begin{frontmatter}

\title{Ferromagnetic transition in a simple variant of the Ising model on multiplex networks}

\author{A. Krawiecki}       

\address{Faculty of Physics,
Warsaw University of Technology, \\
Koszykowa 75, PL-00-662 Warsaw, Poland}

\begin{abstract}
Multiplex networks consist of a fixed set of nodes connected by several sets of 
edges which are generated separately and correspond to different networks ("layers"). 
Here, a simple variant of the Ising model on multiplex networks with two layers is considered, with spins
located in the nodes and edges corresponding to ferromagnetic interactions between them.
Critical temperatures for the ferromagnetic transition are evaluated for the layers in the form of random
Erd\"os-R\'enyi graphs or heterogeneous scale-free networks using the mean-field approximation and the replica method,
from the replica symmetric solution. Both methods require the use of different "partial" magnetizations, 
associated with different layers of the multiplex network, and yield qualitatively similar results.
If the layers are strongly heterogeneous the critical temperature differs noticeably from that 
for the Ising model on a network being a superposition of the two layers, evaluated in the mean-field approximation
neglecting the effect of the underlying multiplex structure on the correlations between the degrees of nodes.
The critical temperature evaluated from the replica symmetric solution depends sensitively on the
correlations between the degrees of nodes in different layers and
shows satisfactory quantitative agreement with that obtained from Monte Carlo simulations.
The critical behavior of the magnetization for the model with strongly heterogeneous layers can depend on the distributions
of the degrees of nodes and is then determined by the properties of the most heterogeneous layer.
\end{abstract}


\end{frontmatter}

\section{Introduction}

Due to the ubiquity and importance of complex networks in many areas of social life, science and technology much research has
been devoted to this topic in the last decades \cite{Albert02,Barabasi16}. Also the physics of interacting systems on complex networks
is a rapidly developing branch of statistical physics \cite{Dorogovtsev08,Barrat08}. Among the latter systems the ferromagnetic (FM) and
spin-glass (SG) transitions in the Ising model was investigated on various complex networks, including heterogeneous scale-free (SF) networks \cite{Barabasi99},
e.g., by means of heterogeneous mean-field (MF) theory \cite{Bianconi02,Leone02,Suchecki06}, the replica method
\cite{Leone02,Kim05}, belief propagation algorithm \cite{Dorogovtsev02,Yoon11} and Monte Carlo (MC) simulations
\cite{Aleksiejuk02,Herrero04,Menche11,Herrero15}. In the context of recent interest in even more complex structures ("networks of networks")
much attention has been devoted to multiplex networks (MNs) which consist of a fixed set of nodes connected by various
sets of edges called layers \cite{Boccaletti14,Lee14,Lee15}. MNs naturally emerge in various social systems (e.g., transportation or communications networks),
and interacting systems on such structures exhibit rich variety of collective behaviors and critical phenomena. For example,
percolation transition \cite{Buldyrev10,Baxter12,Lee12,Min14}, cascading failures \cite{Tan13}, 
threshold cascades \cite{Kim13,Lee14a}, diffusion processes \cite{Gomez13,Sole13}, 
epidemic spreading \cite{Wu16,Zuzek15}, etc., were studied on MNs. In particular, investigation of the Ashkin-Teller model on a MN, 
treated as a model for interacting systems between two species of Ising spins placed on two layers, revealed rich critical behavior
including occurrence of continuous, discontinuous and mixed-order phase transitions, depending on the parameters of the model \cite{Jang15}.

In this paper a simple version of the Ising model on MNs is investigated, 
with spins placed on a fixed set of nodes and with separately generated sets of edges (layers) corresponding to
(in general, different) FM exchange interactions; the layers can have, e.g., a structure of random Erd\"os-R\'enyi (ER) graphs or
heterogeneous SF networks. In contrast with the Ashkin-Teller model on MNs \cite{Jang15} in the Ising model in the absence of the
external field only continuous FM transition is expected to occur. The main aim of this paper is to evaluate the critical temperature for the FM transition 
in the Ising model on MNs with different kinds of layers. In Sec.\ 2 the model is defined. In Sec.\ 3 heterogeneous MF theory for the model
is formulated and the critical temperature in the MF approximation is obtained. In Sec.\ 4 a more rigorous approach based on the replica
method known from the SG theory is developed and again applied to evaluate the critical temperature for the FM transition. In both
approaches analytic calculations are performed for MNs with layers in the form of random ER graphs and heterogeneous SF networks. 
Besides, in Sec.\ 5 the critical behavior of the magnetization is investigated in a certain variant of the model on a MN with heterogeneous layers.
In Sec.\ 6, summary and conclusions are presented.

The main results of the paper are as follows. In the case of layers with high density of connections the critical
temperatures for the FM transition in the Ising model on MNs
obtained from the heterogeneous MF theory and the replica method are close to each other and show
qualitative agreement with results of MC simulations; better quantitative agreement is obtained in the latter case. 
In the case of heterogeneous layers, even if the exchange integrals in all layers are equal, despite the apparent simplicity of the model,
these temperatures can differ noticeably from the critical temperature for the Ising model on a network being a superposition
of the layers (called henceforth a super-network) evaluated using a simple MF approximation neglecting the possible correlations
between the degrees of nodes induced by the underlying multiplex structure. Besides, 
the approach based on the replica method reveals that the critical temperature strongly depends on correlations between the degrees of nodes 
in different layers, which is another example of the
effect of correlated multiplexity observed previously, e.g., in the studies of mutual percolation, robustness 
\cite{Lee12,Min14},  cascading failures \cite{Tan13} and threshold cascades \cite{Kim13} in systems on MNs. 
It also reveals that the critical exponents for the magnetization can differ from their MF values and depend on the properties of
the distribution of the degrees of nodes in the most heterogeneous layer.

\section{The model}

MNs consist of a fixed set of nodes connected by several sets of edges; the set of nodes with each set of edges forms a
network which is called a layer of a MN \cite{Lee14,Lee15}. In the following, for simplicity, MNs with $N$ nodes and only two
layers denoted as $G^{(A)}$, $G^{(B)}$ are considered. The layers (strictly speaking, the sets of edges within each layer)
are generated separately, and, in most cases considered in this paper, independently. As a result, multiple connections between nodes are not
allowed within the same layer, but the same nodes can be connected by multiple edges belonging to different layers.
The nodes $i=1,2,\ldots N$ are characterized by their degrees $k_{i}^{(A)}$,
$k_{i}^{(B)}$ within each layer, i.e., the number of edges attached to them within each layer. The, possibly heterogeneous,
distributions of the degrees of nodes within each layer are denoted as $p_{k^{(A)}}$, $p_{k^{(B)}}$; in the case of independently
generated layers the joint distribution of the degrees of nodes in the MN is 
$p_{k^{(A)},k^{(B)}} = p_{k^{(A)}} p_{k^{(B)}}$. The mean degree of the nodes and the second moment of the
distribution of the degrees of nodes within each layer are denoted as $\langle k^{(A)} \rangle$ ($\langle k^{(B)} \rangle$) and
$\langle k^{(A)2} \rangle$ ($\langle k^{(B)2} \rangle$) for the layer $G^{(A)}$ ($G^{(B)}$), respectively. In this paper only fully overlapping
MNs are considered, with all $N$ nodes belonging to both layers; the case of partly overlapping MNs, with 
 only a fraction of nodes belonging to both layers, is left for future research.

In this paper probably the simplest version of the FM Ising model on a MN with two layers is studied.
The model consists of two-state spins $s_{i}=\pm 1$ located in the nodes $i=1,2\ldots N$ and of edges
within the separately generated layers $G^{(A)}$, $G^{(B)}$ which correspond to exchange interactions with integrals $J^{(A)}>0$, $J^{(B)}>0$, 
respectively. The Hamiltonian of the model is
\begin{equation}
H= - J^{(A)} \sum_{\left( i,j\right) \in G^{(A)}} s_{i}s_{j} - J^{(B)} \sum_{\left( i,j\right) \in G^{(B)}} s_{i}s_{j},
\label{ham}
\end{equation}
where the sums are over all edges belonging to the layer $G^{(A)}$ ($G^{(B)}$). Thus the local field acting on 
the spin in the node $i$ is
\begin{equation}
I_{i}=J^{(A)} \sum_{\left\{j: \left( i,j\right) \in G^{(A)} \right\}} s_{j}  +
J^{(B)} \sum_{\left\{j: \left( i,j\right) \in G^{(B)} \right\}} s_{j},
\label{Ii}
\end{equation}
where the sums are over all nodes $j$ connected to the node $i$ by edges within the layer $G^{(A)}$ ($G^{(B)}$). 
It should be emphasised that in the model under study there is only one spin $s_{i}$ located in each node which interacts with all its neighbors
within all layers. This is in contrast with the related Ashkin-Teller model on a MN with two layers considered in Ref.\ \cite{Jang15},
where spins interacting via exchange interactions within each layer are different, and different spins located in the same node interact 
via additional four-spin interactions. 

The space of parameters of the model is large and comprises $J^{(A)}$, $J^{(B)}$ and characteristics of the distributions 
$p_{k^{(A)}}$, $p_{k^{(B)}}$. Thus, for simplicity, in the following only the Ising model on a MN with
$J^{(A)}=J^{(B)} =J>0$ will be considered. At a first glance this case seems trivial
since the Hamiltonian, Eq.\ (\ref{ham}), is then identical with that for the FM Ising model on a super-network being a superposition of the two layers.
However,  in the most interesting case with heterogeneous layers
any attempt to study the Ising model on the super-network is not straightforward. This is because of the correlations between the 
degrees of nodes $k_{i}=k_{i}^{(A)}+k_{i}^{(B)}$ which inevitably occur due to the separate generation of the layers
(see Sec.\ 3.4.2) and cannot be easily taken into account in the calculations based, e.g., on the heterogeneous MF approximation or the
replica method. However, if they are neglected the obtained critical temperature for the FM transition differs noticeably from that for the Ising
model on a MN.

\section{The mean-field approach}

It is known that first, and in many cases even quantitatively correct approximation for the critical temperature for the FM transition in the Ising model
on heterogeneous networks can be obtained from the heterogeneous MF theory \cite{Bianconi02,Leone02}. Also in the case of MNs the MF approach
was successfully applied, e.g., in the studies of epidemic spreading \cite{Wu16,Zuzek15}. Hence, in this section appropriate theory
is developed for the Ising model on a MN with two separately generated, possibly different layers; examples of such MNs are discussed
in Sec.\ 3.1. For simplicity only the case of independently
generated layers is considered, thus it is possible to assume $p_{k^{(A)},k^{(B)}} = p_{k^{(A)}} p_{k^{(B)}}$ in the calculations. The heterogeneous
MF theory for systems on MNs differs from that for systems on networks since the probabilities that a node is
connected to a node with a given degree must be evaluated separately for each layer. As a result, two magnetization-like "partial" order
parameters, related to the two layers, are necessary to characterize the FM transition in the Ising model, which is shown in Sec.\ 3.2.
General results for the critical temperature in the case of heterogeneous layers are derived in Sec.\ 3.3, and results for the particular cases of
random ER and SF layers are presented in Sec.\ 3.4. Finally, in Sec.\ 3.5 analytic results are compared with those from MC simulations.

\subsection{The network models}

The simplest way to generate a MN with independent layers and with given distributions of the degrees of nodes within layers 
is probably to use the Configuration Model \cite{Newman03} separately and independently for each layer. This method is particularly
useful for generation of heterogeneous SF layers. To generate the first layer $G^{(A)}$,
the algorithm starts with assigning to each node $i$, in a set of $N$ nodes, a degree, i.e., 
a random number $k_{i}^{(A)}$ of ends of edges drawn from a given probability 
distribution $p_{k^{(A)}}$, with $\tilde{m}<  k_{i}^{(A)} < N$ (the minimum degree of node is $\tilde{m}$, and the maximum one $N-1$), with the condition that the sum 
$\sum_{i} k_{i}^{(A)}$ is even. The layer is completed by connecting pairs of the ends of edges chosen uniformly at random to make complete edges, 
respecting the preassigned sequence $k_{i}^{(A)}$ and under the condition that multiple and self-connections are forbidden. Consecutive layers are
generated in a similar way, with the degrees assigned randomly to the nodes from the possibly different probability distributions. Instead, random ER layers
can be generated by selecting randomly and with uniform probability $N\langle k\rangle/2$ pairs of nodes and linking them with edges, 
where $\langle k\rangle$ is the desired mean degree of nodes within the layer \cite{Erdos59}. Using these methods, MNs with
layers with different structure and statistical properties can be easily obtained. 

\subsection{Heterogeneous mean field theory}

There are different ways to derive the MF equations for the order parameter (in general, a sort of magnetization) for the Ising model on,
possibly heterogeneous, networks \cite{Bianconi02,Leone02}. Here we adopt the approach based on
the Master equation for the probability that at time $t$ the system is in the spin configuration $\left( s_{1}, s_{2}, \ldots s_{N}\right)$, 
$s_{i}=\pm 1$,
\begin{eqnarray}
\frac{d}{dt}P\left( s_{1}, s_{2}, \ldots s_{N};t\right)&=&-\sum_{j=1}^{N} w_{j}\left(s_{j}\right)P\left( s_{1}, s_{2}, \ldots, s_{j}, \ldots , s_{N};t\right)
\nonumber \\
&&+ \sum_{j=1}^{N} w_{j}\left(-s_{j}\right)P\left( s_{1}, s_{2}, \ldots,- s_{j}, \ldots , s_{N};t\right),
\label{mastereq}
\end{eqnarray}
where $w_{i}\left( s_{i}\right)$ is 
the transition rate between two spin configurations which differ by a single flip of one spin, e.g., that in the node $i$.
For example, let us assume that the system obeys the Glauber dynamics (used in MC simulations in Sec.\ 3.5)  with
\begin{equation}
w_{i}\left( s_{i}\right) = \frac{1}{2} \left[ 1-s_{i} \tanh \left( \beta I_{i}\right)\right],
\label{Glauber}
\end{equation}
where $\beta=1/T$. Then, multiplying both sides of Eq.\ ({\ref{mastereq}) by $s_{i}$ and performing an ensemble average it is obtained that
\begin{equation}
\frac{d\langle s_{i}\rangle}{dt}= - \langle s_{i}\rangle +\langle \tanh\left(\beta I_{i} \right)\rangle.
\end{equation}
The MF approximation consists in replacing in Eq.\ (\ref{Ii})
\begin{equation}
I_{i} \rightarrow \langle I_{i} \rangle =J \sum_{\left\{j: \left( i,j\right) \in G^{(A)} \right\}} \langle s_{i} \rangle +
J \sum_{\left\{j: \left( i,j\right) \in G^{(B)} \right\}} \langle s_{i} \rangle,
\label{meanI}
\end{equation}
 so that
\begin{equation}
\frac{d\langle s_{i}\rangle}{dt}= - \langle s_{i}\rangle + \tanh\left(\beta \langle I_{i} \rangle\right).
\label{meansi}
\end{equation}

The basic assumption of the heterogeneous MF theory for the Ising model on networks is that the nodes of the network are divided into classes
according to their degrees and that the average values of spins in nodes belonging to the same class are equal. In the case of MNs
the division into classes should be performed with respect to the degrees of nodes within each layer. Thus, for a MN
consisting of two layers $G^{(A)}$, $G^{(B)}$, 
the nodes are divided into classes according to their degrees $\left( k^{(A)}, k^{(B)} \right)$ and it is assumed that the average values of
spins in nodes belonging to each such class is equal to $\langle s_{k^{(A)},k^{(B)} }\rangle$. 
Further analytic results can be obtained if correlations between the degrees of nodes within layers are vanishingly small 
(this is the case of ER layers and SF layers generated from the Configuration Model with $\gamma^{(A)}>3$, $\gamma^{(B)}>3$) or are 
neglected. Then for independent layers the probability that the edge of the layer $G^{(A)}$ attached at one end to the node $i$ is linked 
at the other end to the node with degrees $\left( k^{(A)}, k^{(B)} \right)$ is 
\begin{equation}
\frac{ p_{k^{(A)}} p_{k^{(B)}}k^{(A)}}{\sum_{k^{(A)},k^{(B)}}p_{k^{(A)}} p_{k^{(B)}}k^{(A)}}=
\frac{ p_{k^{(A)}} p_{k^{(B)}}k^{(A)}}{\langle k^{(A)} \rangle},
\label{pkA}
\end{equation}
and similarly for the layer $G^{(B)}$.
Thus, the number of nodes with degrees $\left( k^{(A)}, k^{(B)} \right)$ connected to the node $i$ by edges of the layer $G^{(A)}$ is
\begin{equation}
k_{i}^{(A)} \frac{ p_{k^{(A)}} p_{k^{(B)}}k^{(A)}}{\langle k^{(A)} \rangle},
\end{equation}
and similarly for the layer $G^{(B)}$. Hence, replacing the sums over the indices of nodes by sums over the classes of nodes,
Eq.\ (\ref{meanI}) and ({\ref{meansi}) can be written as
\begin{eqnarray}
\langle I_{i}\rangle &=&
J  k_{i}^{(A)} \sum_{k^{(A)},k^{(B)}} \frac{ p_{k^{(A)}} p_{k^{(B)}}k^{(A)}}{\langle k^{(A)} \rangle}\langle s_{k^{(A)},k^{(B)} }\rangle
\nonumber\\
&&+ J  k_{i}^{(B)} \sum_{k^{(A)},k^{(B)}} \frac{ p_{k^{(A)}} p_{k^{(B)}}k^{(B)}}{\langle k^{(B)} \rangle}\langle s_{k^{(A)},k^{(B)} }\rangle
\nonumber\\
&=&  J \left(  k_{i}^{(A)}\langle S^{(A)} \rangle +  k_{i}^{(B)} \langle S^{(B)} \rangle \right),
\end{eqnarray}
\begin{equation}
\frac{d\langle s_{i}\rangle}{dt}= - \langle s_{i}\rangle + \tanh\left[ J\beta 
\left( k_{i}^{(A)}\langle S^{(A)} \rangle + k_{i}^{(B)} \langle S^{(B)}\rangle \right) \right],
\label{meansi2}
\end{equation}
where the following quantities, referred to as "partial" order parameters, were introduced,
\begin{eqnarray}
\langle S^{(A)} \rangle &\equiv& \frac{1}{N\langle k^{(A)} \rangle} \sum_{i=1}^{N} k_{i}^{(A)} \langle s_{i}\rangle =
\sum_{k^{(A)},k^{(B)}} \frac{ p_{k^{(A)}} p_{k^{(B)}}k^{(A)}}{\langle k^{(A)} \rangle}\langle s_{k^{(A)},k^{(B)} }\rangle
\nonumber\\
\langle S^{(B)} \rangle &\equiv& \frac{1}{N\langle k^{(B)} \rangle} \sum_{i=1}^{N} k_{i}^{(B)} \langle s_{i}\rangle =
\sum_{k^{(A)},k^{(B)}} \frac{ p_{k^{(A)}} p_{k^{(B)}}k^{(B)}}{\langle k^{(B)} \rangle}\langle s_{k^{(A)},k^{(B)} }\rangle.
\label{SASB}
\end{eqnarray}
Multiplying both sides of Eq.\ (\ref{meansi2})
by $\frac{k_{i}^{(A)}}{N \langle k^{(A)} \rangle}$ ($\frac{k_{i}^{(B)}}{N \langle k^{(B)} \rangle}$), performing
the sum over the nodes and replacing it with the sum over the classes of nodes results in the following system of MF equations for the "partial" order parameters
\begin{eqnarray}
\frac{d\langle S^{(A)} \rangle}{dt}&=&- \langle S^{(A)}\rangle + 
\sum_{k^{(A)},k^{(B)}} \frac{ p_{k^{(A)}} p_{k^{(B)}}k^{(A)}}{\langle k^{(A)} \rangle}
\tanh\left[ J\beta \left(  k_{i}^{(A)}\langle S^{(A)} \rangle + k_{i}^{(B)} \langle S^{(B)}\rangle\right) \right] \nonumber\\
\frac{d\langle S^{(B)} \rangle}{dt}&=&- \langle S^{(B)}\rangle + 
\sum_{k^{(A)},k^{(B)}} \frac{ p_{k^{(A)}} p_{k^{(B)}}k^{(B)}}{\langle k^{(B)} \rangle}
\tanh\left[ J\beta \left( k_{i}^{(A)}\langle S^{(A)} \rangle + k_{i}^{(B)} \langle S^{(B)}\rangle\right) \right]. \nonumber\\
&& \label{sAsBMF}
\end{eqnarray}

It should be noted that equations similar to Eq.\ (\ref{sAsBMF}) were obtained for the Ising model on a modular network consisting of two 
heterogeneous networks (modules), with the desity of connections, corresponding to the exchange interactions, within each module much higher 
than that between the modules \cite{Suchecki06,Suchecki09}. The difference is that in the case of modular networks the "partial"
order parameters $\langle S^{(A)} \rangle$, $\langle S^{(B)} \rangle$ are obtained by summing weighted average values of spins
within each module (i.e., the summations are performed over two separate sets of nodes), while in Eq.\ (\ref{SASB}) both summations are over 
the same set of $N$ nodes. A certain degree of separation of nodes in the sums in Eq.\ (\ref{SASB}) can be achieved in partly 
overlapping MNs; investigation of this case is beyond the scope of this paper.

\subsection{General equations for the critical temperature}

Eq.\ (\ref{sAsBMF}) has a fixed point $\left(\langle  S^{(A)} \rangle, \langle S^{(B)} \rangle \right) = (0,0)$
corresponding to the paramagnetic phase. Expanding Eq.\ (\ref{sAsBMF}) in the
vicinity of this fixed point up to linear terms yields
\begin{eqnarray}
\frac{d\langle S^{(A)} \rangle}{dt}&=&\left( -1+\beta J \frac{\langle k^{(A)2}\rangle}{\langle k^{(A)} \rangle}\right)
\langle S^{(A)}\rangle + \beta J \langle k^{(B)} \rangle\langle S^{(B)}\rangle\nonumber \\
\frac{d\langle S^{(B)} \rangle}{dt}&=&\beta J \langle k^{(A)} \rangle  \langle S^{(A)} \rangle
+\left( -1+\beta J \frac{\langle k^{(B)2}\rangle}{\langle k^{(B)} \rangle}\right)\langle S^{(B)}\rangle.
\label{sAsBlinear}
\end{eqnarray}
The paramegnetic fixed point becomes unstable, and the FM phase occurs, if one of the eigenvalues 
of Eq.\ (\ref{sAsBlinear}) crosses zero which takes place if the determinant of the right-hand sides is zero. 
This, in general, leads to two solutions with
\begin{equation}
T_{c\pm}=2J \frac{
 \frac{\langle k^{(A)2}\rangle \langle k^{(B)2}\rangle}{\langle k^{(A)} \rangle \langle k^{(B)} \rangle}
-\langle k^{(A)} \rangle \langle k^{(B)} \rangle }
{ \frac{\langle k^{(A)2}\rangle}{\langle k^{(A)} \rangle}
+\frac{\langle k^{(B)2}\rangle}{\langle k^{(B)} \rangle} \pm \sqrt{\Delta} },
\label{Tcmf}
\end{equation}
where
\begin{displaymath}
\Delta =  \left( \frac{\langle k^{(A)2}\rangle}{\langle k^{(A)} \rangle}
-\frac{\langle k^{(B)2}\rangle}{\langle k^{(B)} \rangle} \right)^{2}
+4 \langle k^{(A)} \rangle \langle k^{(B)} \rangle.
\end{displaymath}
The higher temperature $T_{c-}$ corresponds to the critical temperature for the FM transition, $T_{c}^{MF}=T_{c-}$.
Below $T_{c}^{MF}$ the paramagnetic state is unstable, and the instability at $T=T_{c+} < T_{c}^{MF}$ has no physical meaning.

If the distributions of the degrees of nodes within each layer $G^{(A)}$, $G^{(B)}$ are identical, $p_{k^{(A)}}=p_{k^{(B)}}=p_{k}$, 
and thus $\langle k^{(A)}\rangle =\langle k^{(B)} \rangle =\langle k\rangle$,
$\langle k^{(A)2}\rangle =\langle k^{(B)2} \rangle =\langle k^{2}\rangle$, then
\begin{equation}
T_{c\pm}= J\left( \frac{\langle k^{2}\rangle}{\langle k\rangle}\mp \langle k \rangle\right).
\label{Tcmfident}
\end{equation}
It can be easily verified that in this case the eigenvalue of Eq.\ (\ref{sAsBlinear}) which crosses zero at $T=T_{c}=T_{c-}$ corresponds to the eigenvector with 
$\langle  S^{(A)} \rangle=  \langle S^{(B)} \rangle$, i.e., in fact to the FM phase.

Similar results were obtained for the Ising model on modular networks \cite{Suchecki06,Suchecki09}, where also two values of the critical temperature
were found. The higher value corresponds to the usual FM transition, in which all spins in both modules tend to align in parallel. The lower value
corresponds to the occurrence of another ordered state, in which spins within each module tend to align in parallel, but antiparallel to the spins in the
other module. In the case of small density of connections between spins in different modules this state is a long-living metastable state, and transition
from this state to a stable FM state as the temperature increases can be discontinuous, though only FM intaractions between spins 
are present \cite{Suchecki09}. A similar phenomenon can, in principle, occur in the Ising model on partly overlapping MNs with only a small
fraction of nodes belonging to each layer;  investigation of this case is beyond the scope of this paper.

\subsection{Special cases}

\subsubsection{Random Erd\"os-R\'enyi layers}

If the two layers $G^{(A)}$, $G^{(B)}$ are independently generated random ER graphs 
with mean degrees of nodes $\langle k^{(A)} \rangle$, 
$\langle k^{(B)}\rangle$, respectively, for large $N$ the distributions of the degrees of
nodes can be assumed as Poisson ones, 
$p_{k^{(A)}}=\frac{\langle k^{(A)} \rangle^{k^{(A)}}}{k^{(A)}!}e^{-\langle k^{(A)} \rangle}$,
$p_{k^{(B)}}=\frac{\langle k^{(B)} \rangle^{k^{(B)}}}{k^{(B)}!} e^{-\langle k^{(B)} \rangle}$,
for which $\langle k^{(A)2}\rangle = \langle k^{(A)}\rangle+ \langle k^{(A)}\rangle^{2}$, 
$\langle k^{(B)2}\rangle = \langle k^{(B)}\rangle+ \langle k^{(B)}\rangle^{2}$.
Then the critical temperature, Eq.\ (\ref{Tcmf}), is
\begin{equation}
T_{c}^{MF}=T_{c-}= J \left( 1+ \langle k^{(A)} \rangle + \langle k^{(B)} \rangle \right).
\label{TcmfER}
\end{equation}

Taking into account the way of generation of random ER layers it is easy to see that the resulting MN is also a random ER graph
with the mean degree of nodes $\langle k\rangle = \langle k^{(A)} \rangle + \langle k^{(B)} \rangle$. Thus, it is fully
equivalent to the super-network being a superposition of the two layers, with the degrees of nodes
$k_{i}=k_{i}^{(A)}+k_{i}^{(B)}$ obeying the Poisson distribution $p_{k}=\frac{\langle k \rangle^{k}}{k!} e^{-\langle k \rangle}$ 
with $\langle k\rangle = \langle k^{(A)} \rangle + \langle k^{(B)} \rangle$. Hence, critical temperature for the FM transition
for the Ising model on the super-network is \cite{Leone02,Dorogovtsev02}
\begin{displaymath}
T_{c}^{sup}=J \langle k^{2}\rangle /\langle k\rangle= J\left( 1+\langle k\rangle \right)
= J\left( 1+\langle k^{(A)} \rangle + \langle k^{(B)} \rangle \right) = T_{c}^{FM}
\end{displaymath}
and the results of the MF approach based on the MN and on the corresponding super-network coincide.

\subsubsection{Scale-free layers}

If the two layers $G^{(A)}$, $G^{(B)}$ are independently 
generated SF networks with $p_{k^{(A)}}=\left(\gamma^{(A)}-1\right) \tilde{m}^{\gamma^{(A)}-1}
\left( k^{(A)}\right)^{-\gamma^{(A)}}$, 
 $p_{k^{(B)}}=\left(\gamma^{(B)}-1\right) \tilde{m}^{\gamma^{(B)}-1} \left( k^{(B)}\right)^{-\gamma^{(B)}}$,
with the same minimum node degree $\tilde{m}$
and with the exponents in the power scaling laws of the distributions of the degrees of nodes 
$\gamma^{(A)} >3$, $\gamma^{(B)}>3$, the critical temperature, Eq.\ (\ref{Tcmf}), is
\begin{equation}
T_{c}^{MF}=T_{c -}=2 J \tilde{m} \frac{
\frac{\gamma^{(A)}-2}{\gamma^{(A)}-3}\frac{\gamma^{(B)}-2}{\gamma^{(B)}-3} 
-\frac{\gamma^{(A)}-1}{\gamma^{(A)}-2}\frac{\gamma^{(B)}-1}{\gamma^{(B)}-2}}
{\frac{\gamma^{(A)}-2}{\gamma^{(A)}-3}+ \frac{\gamma^{(B)}-2}{\gamma^{(B)}-3} - \sqrt{\Delta}}, 
\label{TcmfSF}
\end{equation}
with
\begin{displaymath}
\Delta =\left( \frac{\gamma^{(A)}-2}{\gamma^{(A)}-3}- \frac{\gamma^{(B)}-2}{\gamma^{(B)}-3} \right)^{2}
+4 \frac{\gamma^{(A)}-1}{\gamma^{(A)}-2}\frac{\gamma^{(B)}-1}{\gamma^{(B)}-2}. \nonumber
\end{displaymath}

If for at least one layer, say $G^{(A)}$, the distribution of the degrees of nodes has for large $N$ a diverging second moment,
$\langle  k^{(A)2}\rangle  \stackrel{N\rightarrow \infty}{\rightarrow} \infty$, which occurs for SF layers with $2 < \gamma^{(A)} \le 3$, then
expanding in Eq.\ (\ref{Tcmf}) 
\begin{displaymath}
\sqrt{\Delta} \approx \frac{\langle k^{(A)2} \rangle}{\langle k^{(A)}\rangle}
\left(1- \frac{\langle k^{(A)}\rangle}{\langle k^{(A)2}\rangle}
\frac{\langle k^{(B)}\rangle}{\langle k^{(B)2} \rangle} +\ldots \right)
\end{displaymath}
yields
\begin{displaymath}
T_{c}^{MF}=T_{c-}\approx J \frac{\langle k^{(A)2} \rangle}{\langle k^{(A)}\rangle}\stackrel{N\rightarrow \infty}{\rightarrow} \infty,
\end{displaymath}
i.e., there is only crossover (dependent on $N$) temperature for the FM transition and in the thermodynamic limit the system remains in the
FM phase at any temperature. This is the same situation as for the Ising model on any network with a diverging second moment of the
distribution of the degrees of nodes \cite{Leone02,Dorogovtsev02}.

It is interesting to compare the result of Eq.\ (\ref{TcmfSF}) with the critical temperature for the Ising model on a 
super-network in which the degree of each node is a sum of independent degrees 
of this node within each layer, $k_{i}=k_{i}^{(A)}+k_{i}^{(B)}$. The distribution of the degrees of nodes for the
super-network is
\begin{equation}
p_{k}=\left(\gamma^{(A)}-1\right) \tilde{m}^{\gamma^{(A)}-1} \left(\gamma^{(B)}-1\right) \tilde{m}^{\gamma^{(B)}-1} 
\int_{\tilde{m}}^{k-\tilde{m}} \left( k^{(A)}\right)^{-\gamma^{(A)}} \left( k- k^{(A)}\right)^{-\gamma^{(B)}} dk^{(A)}.
\label{pksuper}
\end{equation}
The analytic form of $p_{k}$ is complex \cite{Brennan68,Quang01}, but the distribution can be obtained by evaluating the
integral in Eq.\ (\ref{pksuper}) numerically. In particular, up to leading term, $p_{k}\propto k^{-\gamma_{min}}$ for
$k\ge 2\tilde{m}$, where $\gamma_{min} =\min \left\{ \gamma^{(A)}, \gamma^{(B)} \right\}$. 
Assuming that the possible correlations between the degrees of nodes $k_{i}$ can be neglected the critical temperature
for the FM transition in the Ising model on the super-network is obtained as \cite{Leone02,Dorogovtsev02}
\begin{equation}
T_{c}^{sup} = J \frac{\langle k^2 \rangle}{\langle k\rangle},
\label{Tcmfappr}
\end{equation}
where the averages are taken over $p_{k}$. However, this assumption is equivalent to the one that the probability 
that the edge attached at one end to the node $i$ is linked at the other 
end to the node with degree $k$ is $p_{k}k/\langle k\rangle$, which is obviously not true. In fact,
in the case of a MN with independently generated layers such probabilities should be evaluated separately for each layer and
are given by Eq.\ (\ref{pkA}). 
As a result, noticeable differences between the critical temperatures $T_{c}^{MF}=T_{c-}$ given by Eq.\ (\ref{Tcmf}) and
$T_{c}^{sup}$ given by Eq.\ (\ref{Tcmfappr}) appear (see Sec.\ 3.5).

\subsection{Comparison with Monte Carlo simulations}

\begin{figure}
\centerline{\includegraphics[width=1.0\textwidth]{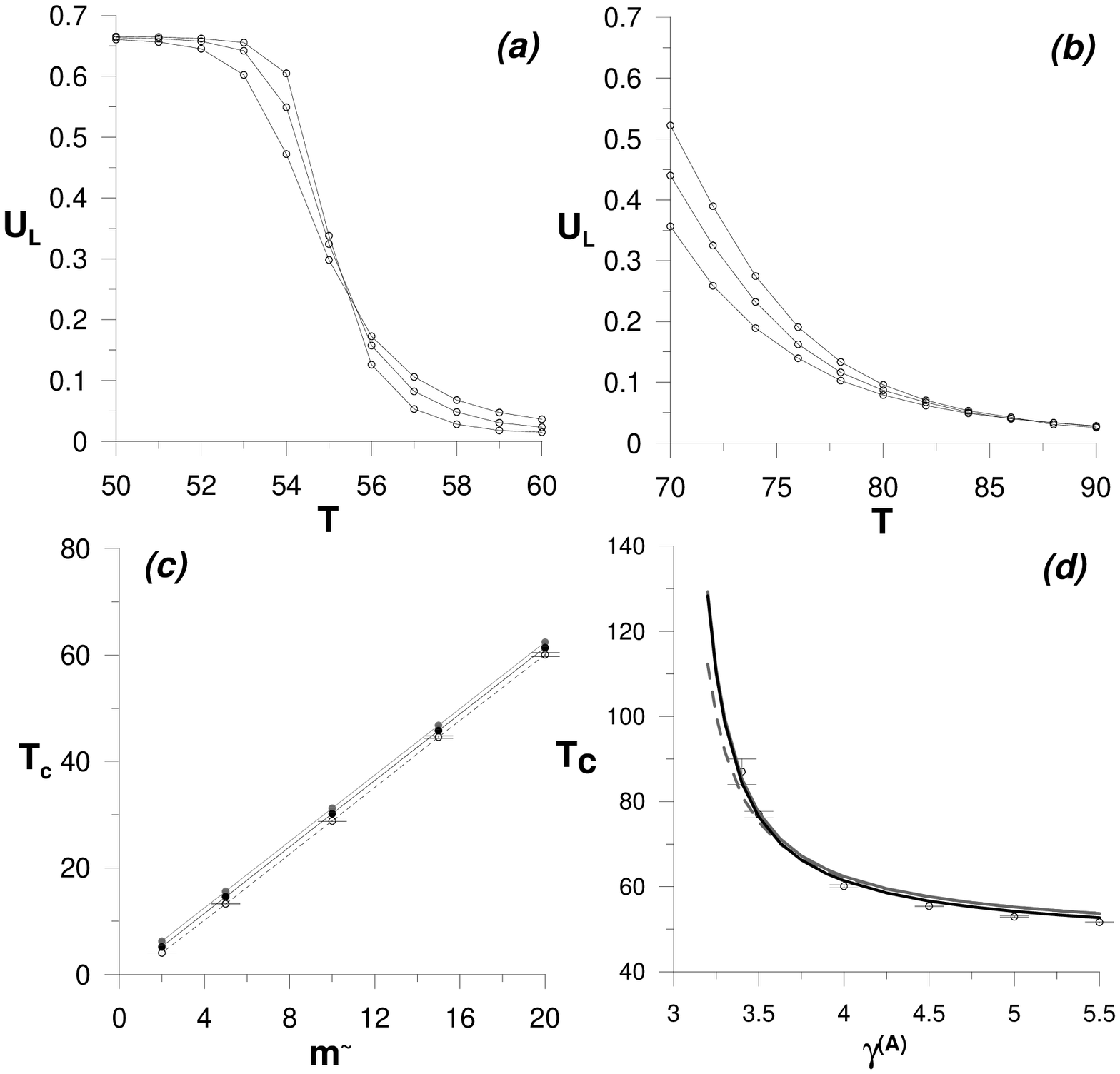}}
\caption{(a) Binder cumulants $U_{L}$ vs.\ $T$ for the Ising model on MNs with SF layers with $J=1$, $\gamma^{(A)} =4.5$,
$\gamma^{(B)} =5.5$ and (from bottom to top for small $T$) $N= 5000$, $10000$, $20000$; (b) as in (a), for
$\gamma^{(A)} =3.4$; (c) Critical temperature vs.\ $\tilde{m}$ for the Ising model on MNs with SF layers
with $J=1$, $\gamma^{(A)} =4.0$, $\gamma^{(B)} =5.5$:
$T_{c}^{MC}$ from MC simulations (circles) and linear least-squares fit
$T_{c}^{MC}=3.12 \tilde{m}+{\rm const}$ (dotted line), 
$T_{c}^{MF}$ from the heterogeneous MF theory, Eq.\ (\ref{TcmfSF}), (gray dots) and
linear least-squares fit $T_{c}^{MF}=3.12 \tilde{m}$ (gray solid line),  
 $T_{c}^{RS}$ from the RS solution, Eq.\ (\ref{Tc3}), (black dots) and
linear least-squares fit $T_{c}^{RS}=3.12 \tilde{m}+{\rm const}$ (black solid line);
(d) Critical temperature vs.\ $\gamma^{(A)}$ for the Ising model on MNs with SF layers with $J=1$, $\gamma^{(B)} =5.5$,
$\tilde{m}=20$:
$T_{c}^{MC}$ from MC simulations (circles),
$T_{c}^{MF}$ from the heterogeneous MF theory, Eq.\ (\ref{TcmfSF}), (gray solid line),  
$T_{c}^{sup}$ for the Ising model on a super-network, from the heterogeneous MF theory, Eq.\ (\ref{Tcmfappr}), (gray dashed line), 
$T_{c}^{RS}$ from the RS solution, Eq.\ (\ref{Tc3}), (black solid line).}
\end{figure}

The Ising model was investigated numerically on MNs with two layers in the form of independent SF networks with $J=1$,
different parameters $\gamma^{(A)}$, $\gamma^{(B)}$, $\tilde{m}$ and numbers of nodes $N$, generated from the 
Configuration Model (Sec.\ 3.1). MC simulations were performed using Glauber dynamics with the single spin-flip
probability $w_{i}\left( s_{i}\right)$ given by Eq.\ (\ref{Glauber}). The critical temperature for the FM transition $T_{c}^{MC}$
was determined from the intersection point of the Binder cumulants $U_{L}$ vs.\ $T$ for different $N$ \cite{Binder97},
\begin{equation}
U_{L}=\left[ 1-\frac{\langle M^{4} \rangle_{t}}{3\langle M^{2}\rangle_{t}^{2}}\right]_{av},
\label{UL}
\end{equation}
where $M=N^{-1}\sum_{i=1}^{N} s_{i}$ is the usual magnetization, $\langle \cdot \rangle_{t}$ denotes the time average
for the simulation of the Ising model on a particular MN, and $\left[ \cdot \right]_{av}$ denotes the average over random realizations of the 
MN with given $N$ and distributions of the degrees of nodes $p_{k^{(A)}}$, $p_{k^{(B)}}$. Exemplary curves $U_{L}$ vs.\
$T$ are shown in Fig.\ 1(a,b) for fixed $\gamma^{(B)}=5.5$ and different $\gamma^{(A)}$. For $\gamma^{(A)} \ge 3.5$
the intersection point of the cumulants for different $N$ can be determined with high accuracy (Fig.\ 1(a)). For $\gamma^{(A)}< 3.5$
the intersection, if any, occurs in the region of temperatures where the curves are flat and close to zero, and $T_{c}^{MC}$
is either determined with relatively high error or cannot be determined at all (Fig.\ 1(b)). 

Comparison of the critical temperatures obtained from the MC simulations and from the heterogeneous MF theory, Eq.\ (\ref{TcmfSF}),
for the Ising model on MNs with SF layers with $J=1$ and with fixed $\gamma^{(A)}>3$, $\gamma^{(B)}>3$ and different $\tilde{m}$
(Fig.\ 1(c)) shows that $T_{c}^{MF}$ is systematically higher than $T_{c}^{MC}$.
The first source of this discrepancy is the approximate MF character of Eq.\ (\ref{TcmfSF}) as well as the more general
Eq.\ (\ref{Tcmf}). The second one originates from the fact that in the analytic calculations leading to Eq.\ (\ref{TcmfSF})
the distributions $p_{k^{(A)}}$, $p_{k^{(B)}}$ of the degrees of nodes within the layers were assumed continuous, with $k^{(A)} \ge \tilde{m}$,
$k^{(B)} \ge \tilde{m}$, while in the layers generated using the Configuration Model only integer values of the degrees of nodes 
are allowed and there are upper constraints for the maximum degrees, $k^{(A)} < N$, $k^{(B)} < N$. 
Nevertheless, the relative difference between $T_{c}^{MF}$ and $T_{c}^{MC}$ is not large and
decreases with $\tilde{m}$, and thus with $\langle k^{(A)}\rangle$, $\langle k^{(B)}\rangle$, as expected for the MF theory.
Moreover, both critical temperatures show the same linear dependence on $\tilde{m}$ predicted by Eq.\ (\ref{TcmfSF}).
These results are qualitatively similar to those obtained from the heterogeneous MF theory and MC simulations of the Ising model on
SF networks \cite{Herrero04}.

In Fig.\ 1(d) the critical temperatures for the Ising model on MNs with SF layers with $J=1$, fixed $\gamma^{(B)}=5.5$ and
high $\tilde{m}=20$ and with different $\gamma^{(A)}>3$ are shown. It can be seen that $T_{c}^{MF}$ diverges as
$\gamma^{(A)}\rightarrow 3$, and thus $\langle k^{(A)2} \rangle \rightarrow\infty$, as expected. For $\gamma^{(A)} \ge 4$ 
the MF critical temperature overestimates that obtained from MC simulations, as discussed above. On the other hand, for
$\gamma^{(A)}\le 3.5$ the diverging $T_{c}^{MF}$ shows even quantitative agreement with $T_{c}^{MC}$;
unfortunately, as mentioned above, for $\gamma^{(A)}\rightarrow 3$ the critical temperature cannot be accurately determined from 
MC simulations and comparison between numerical and analytic results is dubious.

In Fig.\ 1(d) the critical temperatures $T_{c}^{sup}$, Eq.\ (\ref{Tcmfappr}), are also shown 
for the Ising model on a super-network being a superposition of two SF layers
with $J=1$, fixed $\gamma^{(B)}=5.5$ and high $\tilde{m}=20$ and with different $\gamma^{(A)}>3$.
It can be seen that $T_{c}^{sup}$ is almost equal to $T_{c}^{MF}$ for $\gamma^{(A)}\approx \gamma^{(B)}$
(in fact, for $\gamma^{(A)}=\gamma^{(B)}$ there is $T_{c}^{sup}=T_{c}^{MF}$, and the value of the MF 
critical temperature is given by Eq.\ (\ref{Tcmfident})). However, $T_{c}^{sup}$ deviates from $T_{c}^{MF}$ 
(is systematically smaller) as $\gamma^{(A)}\rightarrow 3$; as a result, $T_{c}^{sup}$ in this range of parameters 
underestimates the critical temperature $T_{c}^{MC}$ obtained from MC simulations. This shows that 
neglecting completely the underlying multiplex structure of the super-network leads to approximate results
for the critical temperature which can differ noticeably from $T_{c}^{FM}$ obtained from the MF theory
for the Ising model on a heterogeneous MN. In fact, similar difference
was observed if only one layer was a SF network, and the other one was, e.g., a random ER or random regular graph.

\section{The approach using the replica method}

In this section the simple version of the FM Ising model on MNs is investigated using a method typical for the SG theory,
namely the replica method and, in particular, the replica-symmetric solution \cite{Mezard87,Nishimori01}. 
Assuming purely FM interactions between all pairs of spins within all layers
and using this method it is possible to evaluate critical temperature for the FM transition \cite{Leone02,Kim05}; in the case of the Ising model
on heterogeneous networks such critical temperature agrees quantitatively with that obtained from 
MC simulations. In Sec.\ 4.1 the class of models of MNs is defined for which the critical temperature is 
evaluated from the replica method. These models are based on the so-called static model \cite{Goh01,Lee04}: each layer of the MN
is generated by first assigning a weight to each of $N$ nodes and then connecting the nodes with edges taking into 
account the prescribed weights; weights associated with the same node can be different for different layers. FM
and SG transitions in the Ising model on random ER graphs and heterogeneous networks obtained from the static model were
investigated in Ref.\ \cite{Viana85} and \cite{Kim05}, respectively, and the considerations in Sec.\ 4.2 and 4.3 below are a straightforward generalization
of these studies to the case of MNs. In Sec.\ 4.4 and 4.5 results for the FM critical temperature are presented for
various MNs obtained from the static model
and their generalization is proposed to the case of MNs with layers in the form of general heterogeneous networks.
In Sec.\ 4.6 results from the replica method are compared with those from the MF theory,
Sec.\ 3, and with MC simulations.

\subsection{The network models}

Using the static model the (possibly heterogeneous) networks with a fixed number of nodes $N$ and desired distributions of the degrees of nodes can be generated as follows
\cite{Goh01,Lee04}. First, a weight $v_{i}$ is assigned to each node so that the condition $\sum_{i=1}^{N} v_{i}=1$ was fulfilled. Then, 
nodes are linked with edges in accordance with the prescribed sequence of weights, by selecting
a pair of nodes $i$, $j$ ($i\neq j$) with probablities $v_{i}$, $v_{j}$, respectively, linking them with an edge and repeating this process $NK/2$ times. In this way,
a network is obtained with the probability that the nodes $i$, $j$ are linked by an edge $f_{ij}\approx NK v_{i}v_{j}$,
with the mean degree of nodes $\langle k\rangle =K$, 
and with the distribution of the degrees of nodes depending on the choice of the weights. In particular, random ER graph is obtained if $v_{i}=1/N$ 
is assumed for all $i$. 
For a sequence $v_{i}=i^{-\mu}/\zeta_{N}( \mu)$ associated with the nodes $i=1,2,\ldots N$, where $0< \mu <1$ and $\zeta_{N}( \mu ) \approx N^{1-\mu}/(1-\mu)$,
SF network is obtained with the distribution of the degrees of nodes $p_{k}\propto k^{-\gamma}$, $\gamma = 1 +1/\mu$. 
In an ensemble of networks generated from the static model the mean degree of a given node $i$ is $\langle k_{i}\rangle =v_{i}$. 

The MN with a fixed set of nodes and two layers $G^{(A)}$, $G^{(B)}$ is generated by associating weights $v_{i}^{(A)}$, $v_{i}^{(B)}$ with the nodes 
separately to generate each layer. In this way, the layers can have different distributions of the degerees of nodes $p_{k^{(A)}}$, $p_{k^{(B)}}$. Let us note that
the numbering of nodes $i=1,2,\ldots N$ while generating each layer can be assumed the same or different. In the case of random ER layers this distinction is unimportant, however,
in the case of SF layers it can introduce correlations between the two sequences of weights $v_{i}^{(A)}$, $v_{i}^{(B)}$, $i=1,2,\ldots N$, where now and 
henceforth $i$ denotes the index of the node in a MN, common for all layers. In particular, a MN with independent
layers is obtained by randomly and independently associating weights from the two appropriate sets of weights with the nodes and then linking them with
edges according to the prescribed sequence of weights within each layer.

\subsection{Evaluation of the free energy}

In order to go beyond the MF approximation the thermodynamic potentials for the Ising model should be evaluated on a statistical ensemble of 
MNs generated according to a given rule. Hence, the free energy is 
$-\beta F=\left[ \ln Z\right]_{av}$, where $Z$ is the partition function for the Ising model on a particular MN, 
and  the average $\left[ \cdot \right]_{av}$ is taken over all possible random realizations of a MN of a given kind
(since all non-zero exchange integrals are assumed as $J_{ij}^{(A)}=J_{ij}^{(B)} =J>0$, 
there is no usual averaging over the quenched disorder of the exchange interactions). 
In the framework of the replica method the free energy is formally evaluated as 
$-\beta F =\lim_{n\rightarrow 0} \left\{ \left[ Z^{n}\right]_{av} -1\right\}/n$. The 
average of the $n$-th power of the partition function is
\begin{equation}
\left[ Z^{n}\right]_{av} = {\rm Tr}_{\{s^{\alpha}\}}
\left[ \exp\left( \beta J \sum_{\left( i,j\right)\in G^{(A)}} \sum_{\alpha =1}^{n}s_{i}^{\alpha}s_{j}^{\alpha}\right)
\exp\left( \beta J \sum_{\left( i,j\right)\in G^{(B)}} \sum_{\alpha =1}^{n}s_{i}^{\alpha}s_{j}^{\alpha}\right) \right]_{av},
\label{Zn}
\end{equation}
i.e., it is the average of a product of $n$ partition functions for non-interacting replicas (copies) of the system,
the trace ${\rm T}_{\left\{ s^{\alpha}\right\}}$ is taken over all replicated spins $s_{i}^{\alpha} =\pm 1$, and $\alpha =1,2\ldots n$ is the replica
index.

It should be emphasised that generation of a MN takes place in two stages: first, in which the weights
$v_{i}^{(A)}$, $v_{i}^{(B)}$ are separately assigned to the nodes $i=1,2,\ldots N$, and second, in which the nodes are connected
with edges taking into account the prescribed weights within each layer. Thus, in principle, the average in Eq.\ (\ref{Zn}) should be
taken over all possible realizations of the two above-mentioned independent random processes. However, at the first stage 
the two sequences of weights $v_{i}^{(A)}$, $v_{i}^{(B)}$, $i=1,2,\ldots N$ can be assigned to the nodes independently, or 
certain correlations between them can be present. Then, the degrees
of nodes $k_{i}^{(A)}$, $k_{i}^{(B)}$ within each layer can be also, on average,
independent or correlated which can significantly
affect the critical temperature for the FM transition, and averaging over all pairs of sequences of weights in Eq.\ (\ref{Zn}) 
would hide the latter dependence. An important role of correlations between the degrees of nodes in layers of MNs
was also emphasised in other phenomena, as percolation, cascading failures and threshold cascades \cite{Lee12,Min14,Tan13,Kim13}.
Thus, it is reasonable to consider separately classes of MNs characterized
by given pairs of sequences of weights $v_{i}^{(A)}$, $v_{i}^{(B)}$,  $i=1,2,\ldots N$. Then the average in
Eq.\ (\ref{Zn}) is evaluated separately for each class, and is taken over all possible realizations of the two layers by connecting the
nodes with edges according to the weights $v_{i}^{(A)}$, $v_{i}^{(B)}$, $i=1,2,\ldots N$ characterizing this class. If
necessary, a sort of further averaging over different classes of MNs (e.g., over all classes with the same 
correlation coefficient between the two sequences of weights $v_{i}^{(A)}$, $v_{i}^{(B)}$, $i=1,2,\ldots N$) can be
performed by replacing the sums over $N$ nodes by their expected values in the resulting expressions for the critical 
temperature.

For a class of MNs with fixed assignment of the weights $v_{i}^{(A)}$, $v_{i}^{(B)}$ to the nodes, the average in Eq.\ (\ref{Zn})
can be taken independently over all realizations of the layers $G^{(A)}$ and $G^{(B)}$ in accordance with these weights.
Denoting the respective averages by $\left[ \cdot \right]_{av}^{(A)}$, $\left[ \cdot \right]_{av}^{(B)}$ it is obtained that
\begin{equation}
\left[ Z^{n}\right]_{av}= {\rm Tr}_{\{s^{\alpha}\}}
\left[ \exp\left( \beta J \sum_{\left( i,j\right)\in G^{(A)}} \sum_{\alpha =1}^{n}s_{i}^{\alpha}s_{j}^{\alpha}\right) \right]_{av}^{(A)}
\left[ \exp\left( \beta J \sum_{\left( i,j\right)\in G^{(B)}} \sum_{\alpha =1}^{n}s_{i}^{\alpha}s_{j}^{\alpha}\right) \right]_{av}^{(B)}.
\label{Zn1}
\end{equation}
The two factors can be evaluated as in Ref.\ \cite{Kim05},
\begin{eqnarray}
&& \left[ \exp\left( \beta J \sum_{\left( i,j\right)\in G^{(A)}} \sum_{\alpha =1}^{n}s_{i}^{\alpha}s_{j}^{\alpha}\right) \right]_{av}^{(A)} =  \nonumber\\
&& \prod_{i<j} \left[ \left( 1-f_{ij}^{(A)} \right) + f_{ij}^{(A)} \exp\left( \beta J \sum_{\alpha =1}^{n}s_{i}^{\alpha}s_{j}^{\alpha}\right) \right]  = \nonumber\\
&& \exp \left\{ \sum_{i<j} \ln \left[ 1+f_{ij}^{(A)} \left( \exp\left( \beta J  \sum_{\alpha =1}^{n}s_{i}^{\alpha}s_{j}^{\alpha}\right) -1\right)\right] \right\} 
\approx \nonumber\\
&& \exp\left[ \sum_{i<j} NK^{(A)} v_{i}^{(A)}v_{j}^{(A)}\left( \exp\left( \beta J \sum_{\alpha =1}^{n}s_{i}^{\alpha}s_{j}^{\alpha}\right) -1\right) \right],
\label{Zn2}
\end{eqnarray}
and similarly for the average $\left[ \cdot \right]_{av}^{(B)}$. Then, since $s_{i}^{\alpha}s_{j}^{\alpha} = \pm 1$, the relation
\begin{equation}
\exp\left( \beta J \sum_{\alpha =1}^{n}s_{i}^{\alpha}s_{j}^{\alpha}\right) = \prod_{\alpha} \cosh \beta J \left( 1+ s_{i}^{\alpha}s_{j}^{\alpha} 
\tanh\beta J \right)
\end{equation}
can be used in Eq.\ (\ref{Zn2}), which yields
\begin{eqnarray}
&& \left[ \exp\left( \beta J \sum_{\left( i,j\right)\in G^{(A)}} \sum_{\alpha =1}^{n}s_{i}^{\alpha}s_{j}^{\alpha}\right) \right]_{av}^{(A)} \propto  \nonumber\\
&& \exp\left[ \sum_{i<j} NK^{(A)} v_{i}^{(A)}v_{j}^{(A)} \left( {\bf T}_{1} \sum_{\alpha} s_{i}^{\alpha}s_{j}^{\alpha} +
{\bf T}_{2} \sum_{\alpha < \beta} s_{i}^{\alpha}s_{i}^{\beta} s_{j}^{\alpha}s_{j}^{\beta} +\ldots \right) \right],
\label{Zn4}
\end{eqnarray}
where ${\bf T}_{1}=\cosh^{n}\beta J \tanh\beta J$, ${\bf T}_{2}=\cosh^{n}\beta J \tanh^{2}\beta J$, etc; 
similar expansion can be obtained for the average $\left[ \cdot \right]_{av}^{(B)}$.
Finally, after applying the Hubbard-Stratonovich identity to the expressions of the form (\ref{Zn4}), separately for the two averages 
$\left[ \cdot \right]_{av}^{(A)}$, $\left[ \cdot \right]_{av}^{(B)}$, and grouping terms connected with the same nodes $i$ it is obtained that
\begin{eqnarray}
&& \left[ Z^{n}\right]_{av} = \nonumber \\
&& \int dq_{\alpha}^{(A)} \int dq_{\alpha \beta}^{(A)}\ldots \int dq_{\alpha}^{(B)} \int dq_{\alpha\beta }^{(B)}\ldots 
\exp \left[ -Nn\beta f \left( q_{\alpha}^{(A)}, q_{\alpha\beta}^{(A)},\ldots q_{\alpha}^{(B)}, q_{\alpha\beta }^{(B)} \ldots \right) \right] \nonumber \\
&& \equiv \int \ d{\bf q} \exp \left[ -Nn\beta f({\bf q}) \right],
\label{Zn6}
\end{eqnarray}
with
\begin{eqnarray}
n\beta  f({\bf q})& =& \frac{K^{(A)} {\bf T}_{1}}{2} \sum_{\alpha} q_{\alpha}^{(A)2} + \frac{K^{(B)} {\bf T}_{1}}{2} \sum_{\alpha} q_{\alpha}^{(B)2} 
\nonumber\\
&+& \frac{K^{(A)} {\bf T}_{2}}{2} \sum_{\alpha < \beta} q_{\alpha\beta}^{(A)2} + 
\frac{K^{(B)} {\bf T}_{2}}{2} \sum_{\alpha< \beta} q_{\alpha\beta}^{(B)2} +\ldots \nonumber \\
&-& \frac{1}{N} \sum_{i} \ln {\rm Tr}_{\left\{ s_{i}^{\alpha}\right\}} \exp\left(  X_{i}^{(A)} +X_{i}^{(B)} \right),
\label{Zn5}
\end{eqnarray}
where ${\rm Tr}_{\left\{ s_{i}^{\alpha}\right\}}$ is the trace over the replicated spins at node $i$, and
\begin{equation}
X_{i}^{(A)} = NK^{(A)}{\bf T}_{1} v_{i}^{(A)} \sum_{\alpha}q_{\alpha}^{(A)} s_{i}^{\alpha} +
 NK^{(A)}{\bf T}_{2} v_{i}^{(A)} \sum_{\alpha < \beta}q_{\alpha\beta}^{(A)} s_{i}^{\alpha} s_{i}^{\beta} +\ldots,
\end{equation}
and similarly for $X_{i}^{(B)}$. 

The elements of a set $\{\bf q\}$, $q_{\alpha}^{(A)}, q_{\alpha\beta}^{(A)}, \ldots, q_{\alpha}^{(B)}, q_{\alpha\beta}^{(B)},
\ldots $ form in a natural way two subsets of an infinite set of the order parameters associated with the two layers of the MN
$G^{(A)}$, $G^{(B)}$. The first two order parameters, called magnetizations for convenience,
\begin{equation}
q_{\alpha}^{(A)} = \sum_{i}v_{i}^{(A)} \overline{s_{i}^{\alpha}}, \; \; q_{\alpha}^{(B)} = \sum_{i}v_{i}^{(B)} \overline{s_{i}^{\alpha}}, 
\label{ordpar}
\end{equation}
where the averages are evaluated as 
\begin{displaymath}
\overline{s_{i}^{\alpha}} = \frac{{\rm Tr}_{\left\{ s_{i}^{\alpha}\right\}} s_{i}^{\alpha}  \exp\left(  X_{i}^{(A)} +X_{i}^{(B)} \right)}
{{\rm Tr}_{\left\{ s_{i}^{\alpha}\right\}}  \exp\left(  X_{i}^{(A)} +X_{i}^{(B)} \right)},
\end{displaymath}
resemble the "partial" order parameters $\langle S^{(A)}\rangle$, $\langle S^{(B)}\rangle$ of the MF approach in Sec.\ 3, 
Eq.\ (\ref{SASB}), with the average spin values 
weighted by $v_{i}^{(A)}$, $v_{i}^{(B)}$ rather than directly by the degrees of nodes $k_{i}^{(A)}$,
$k_{i}^{(B)}$ within each layer. All remaining order parameters are unimportant for the evaluation of the 
critical temperature for the FM Ising model.

\subsection{The replica symmetric free energy}

The simplest replica symmetric (RS) solution for the order parameters is obtained under the assumption that spins with different replica index
are indistinguishable. Thus in the case of the Ising model on a heterogeneous network the solution in the form
$q_{\alpha}=m$, $q_{\alpha \beta}=q$, etc., for $\alpha, \beta =1,2\ldots n$, etc., is looked for \cite{Kim05}.
However, in the case of a MN in general different weights $v_{i}^{(A)}$, $v_{i}^{(B)}$, $i=1,2\ldots N$  occur 
in the sums over the indices of nodes, Eq.\  (\ref{ordpar}), definining the two sets of the order parameters $q_{\alpha}^{(A)}, q_{\alpha \beta}^{(A)}, \ldots$ and 
$q_{\alpha}^{(B)}, q_{\alpha \beta}^{(B)}, \ldots$. Thus, it is more natural to look for the solution in the form
$q_{\alpha}^{(A)} =m^{(A)}$, $q_{\alpha \beta}^{(A)}=q^{(A)}$, etc., and
$q_{\alpha}^{(B)} =m^{(B)}$, $q_{\alpha \beta}^{(B)}=q^{(B)}$, etc., for $\alpha, \beta =1,2\ldots n$, etc., where, in general,
$m^{(A)}\neq m^{(B)}$, $q^{(A)}\neq q^{(B)}$, etc. It should be emphasised that this assumption has nothing to do with breaking the
RS and follows simply from the existence of the two sets of the order parameters, associated with the layers $G^{(A)}$, $G^{(B)}$ of the
MN.

Since at the present stage of research we are mainly interested in the evaluation of the critical temperature for the Ising model with purely
FM interactions it is enough to retain only the terms containing magnetizations in the free energy, Eq.\ (\ref{Zn5}), and truncate it at the order $m^2$.
Assuming the above-mentioned form of the RS solution it is obtained that
\begin{equation}
n\beta f\left( m^{(A)},m^{(B)}\right)= 
\frac{K^{(A)} {\bf T}_{1}}{2}nm^{(A)2}+\frac{K^{(B)} {\bf T}_{1}}{2}nm^{(B)2} +\ldots  - \frac{1}{N} \sum_{i} \ln {\cal Z}_{i},
\label{nBf}
\end{equation}
where
\begin{equation}
{\cal Z}_{i}= {\rm Tr}_{\left\{ s_{i}^{\alpha}\right\}}\exp \left( \eta_{i} \sum_{\alpha} s_{i}^{\alpha} + \ldots \right)\approx \left(2 \cosh \eta_{i}\right)^{n},
\label{Zi}
\end{equation}
with
\begin{equation}
\eta_{i} = N {\bf T}_{1} \left( K^{(A)} v_{i}^{(A)} m^{(A)} + K^{(B)} v_{i}^{(B)} m^{(B)} \right).
\end{equation}
Thus, the important part of the free energy is
\begin{equation}
\beta f\left( m^{(A)},m^{(B)}\right)= \frac{K^{(A)} {\bf T}_{1}}{2}m^{(A)2}+\frac{K^{(B)} {\bf T}_{1}}{2}m^{(B)2} -
\frac{1}{N}\sum_{i} \ln \left(2 \cosh \eta_{i}\right)
\label{Bf}
\end{equation}
with ${\bf T}_{1}=\tanh \beta$ in the limit  $n\rightarrow 0$.

With $\beta f({\bf q})$ given by Eq.\ (\ref{Zn5}) the integral in Eq.\ (\ref{Zn6}) can be evaluated using the saddle point method. For this
purpose, assuming the RS solution, the minimum of the function $f\left( m^{(A)},m^{(B)}\right)$ should be found, and the necessary condition for
the existence of the extremum leads to the following set of self-consistent equations for the magnetizations $m^{(A)}$, $m^{(B)}$,
\begin{eqnarray}
\frac{\partial f}{\partial m^{(A)}} &=&m^{(A)}-\sum_{i=1}^{N}v_{i}^{(A)} \tanh \eta_{i}=0, \nonumber\\
\frac{\partial f}{\partial m^{(B)}} &=&m^{(B)}-\sum_{i=1}^{N}v_{i}^{(B)} \tanh \eta_{i}=0.
\label{system}
\end{eqnarray}

\subsection{General equations for the critical temperature}

For small $m^{(A)}$, $m^{(B)}$ after expanding the logarithm in Eq.\ (\ref{Bf}) the free
energy can be written as
\begin{eqnarray}
&& \beta f\left( m^{(A)},m^{(B)} \right)= \nonumber\\
&& \frac{K^{(A)} {\bf T}_{1}}{2}m^{(A)2}+\frac{K^{(B)} {\bf T}_{1}}{2}m^{(B)2} 
- \frac{1}{2}N \sum_{i=1}^{N} {\bf T}_{1}^{2} \left( K^{(A)} v_{i}^{(A)} m^{(A)} + K^{(B)} v_{i}^{(B)} m^{(B)} \right)^{2}. \nonumber\\
\label{Bflinear}
\end{eqnarray}
Then equating the derivatives to zero in Eq.\ (\ref{system}) leads to the following system of linear equations for $m^{(A)}$, $m^{(B)}$,
\begin{eqnarray}
\left( 1-NK^{(A)} {\bf T}_{1} \sum_{i=1}^{N} v_{i}^{(A)2} \right) m^{(A)} 
-NK^{(B)}{\bf T}_{1}\left( \sum_{i=1}^{N} v_{i}^{(A)}v_{i}^{(B)}\right) m^{(B)} &=&0 \nonumber\\
-NK^{(A)}{\bf T}_{1}\left( \sum_{i=1}^{N} v_{i}^{(A)}v_{i}^{(B)}\right) m^{(A)}
+\left( 1-NK^{(B)} {\bf T}_{1} \sum_{i=1}^{N} v_{i}^{(B)2} \right) m^{(B)} &=& 0.\nonumber\\
&& \label{mAmB}
\end{eqnarray}
Non-zero solutions of the system of Eq.\ (\ref{mAmB}) exist if the determinant is zero; from this condition the critical temperature for the FM
transition from the RS solution, denoted as $T_{c}^{RS}$, 
can be obtained. It can be seen that the critical temperature depends on the correlation $\rho=\sum_{i=1}^{N} v_{i}^{(A)}v_{i}^{(B)}$
between the two sequences of weights 
$v_{i}^{(A)}$, $v_{i}^{(B)}$, $i=1,2,\ldots N$, assigned to the nodes during the generation of the two layers of the MN.
In the next section the critical temperature is evaluated directly for several cases, e.g., for the Ising model on MNs
with layers in the form of random ER graphs or SF networks.

\subsection{Special cases}

\subsubsection{Random Erd\"os-R\'enyi layers}

If the two layers of the MN are random ER graphs with mean degrees of nodes
$K^{(A)}$, $K^{(B)}$ all weights are equal,
$v_{i}^{(A)}=v_{i}^{(B)}=1/N$, and the critical temperature for the FM transition is
\begin{equation}
T_{c}^{RS}=J {\rm atanh}^{-1} \left(\frac{1}{K^{(A)}+K^{(B)}}\right).
\end{equation}
This is of course the critical temperature for the FM transition in the Ising model on a random ER graph with the mean degree of
nodes $K^{(A)}+K^{(B)}$. For large $K^{(A)}$, $K^{(B)}$ this result agrees with that from the MF approximation, Eq.\ (\ref{TcmfER}), 
and is obvious since, as mentioned in Sec.\ 3.4.1, the two-layer MN is equivalent to a random ER graph with the mean degree of nodes
equal to the sum of mean degrees of nodes within each layer.

\subsubsection{Independent scale-free layers}

Let us consider the case when the two layers of the MN are SF networks obtained from the static models with parameters
$\mu^{(A)}$, $\mu^{(B)}$, with the 
distributions of the degrees of nodes obeying power scaling laws, 
$p_{k^{(A)}}\propto k_{A}^{-\gamma^{(A)}}$, $\gamma^{(A)}=1+1/\mu^{(A)}$
$p_{k^{(B)}}\propto k_{B}^{-\gamma^{(B)}}$, $\gamma^{(B)}=1+1/\mu^{(B)}$.
First let us focus on the case when the layers are generated 
independently.
To generate the layer $G^{(A)}$, the weights $v_{i}^{(A)}$ are randomly assigned to the nodes from the set of
weights $v_{j}=j^{-\mu^{(A)}}/\zeta_{N}(\mu^{(A)})$, $j=1,2\ldots N$, and then the nodes are connected with $NK^{(A)}/2$ edges in
accordance with the prescribed sequence of weights $v_{i}^{(A)}$. Next, the same procedure is repeated for the layer $G^{(B)}$,
with the weights $v_{i}^{(B)}$ randomly assigned to the nodes from the set of weights $v_{l}=l^{-\mu^{(B)}}/\zeta_{N}(\mu^{(B)})$, $l=1,2\ldots N$.
This means that the two sequences of weights $v_{i}^{(A)}$, $v_{i}^{(B)}$, $i=1,2,\ldots N$ are independent.
As a result, in Eq.\ (\ref{mAmB}) the sum over the products of weights can be approximated by its expected value,
\begin{eqnarray}
\rho &=& N\sum_{i=1}^{N} v_{i}^{(A)}v_{i}^{(B)} \approx N \langle \sum_{i=1}^{N} v_{i}^{(A)}v_{i}^{(B)} \rangle =
N \sum_{i=1}^{N} \langle v_{i}^{(A)}v_{i}^{(B)} \rangle \nonumber\\
&&  =N \sum_{i=1}^{N}\sum_{j=1,l=1}^{N}\frac{1}{N^{2}} v_{j}v_{l}
= N \frac{1}{N^{2}}\sum_{i=1}^{N} \left( \sum_{j=1}^{N} v_{j}\right) \left(\sum_{l=1}^{N} v_{l}\right) = 1.
\label{expval}
\end{eqnarray}
This approximation is valid in typical cases of MNs with independently generated layers, and is applied instead of
averaging the partition function in Eq.\ (\ref{Zn}) over a class of MNs with mutually independent sequences of weights
assigned to nodes when generating different layers.
Besides, for $\mu^{(A)}<1/2$ ($\gamma^{(A)}>3$), $\mu^{(B)}<1/2$ ($\gamma^{(B)}>3$) there is
\begin{equation}
N\sum_{i=1}^{N}v_{i}^{(A)2}= N\sum_{j=1}^{N}v_{j}^{2}\approx \frac{\left(1-\mu^{(A)}\right)^2}{1-2\mu^{(A)}}
=\frac{\left( \gamma^{(A)}-2 \right)^{2}}{\left( \gamma^{(A)}-1\right) \left( \gamma^{(A)}-3\right)},
\label{sump2}
\end{equation}
and similarly for $\sum_{i=1}^{N} v_{i}^{(B)2}$. 
Equating to zero the determinant of Eq.\ (\ref{mAmB}), 
as in the case of the critical temperature for the FM transition obtained in the MF approximation (Sec.\ 3.3), leads in general to two solutions
with respect to the temperature, of which that with higher value corresponds to the critical temperature for the FM
transition,
\begin{equation}
T_{c}^{RS}=J {\rm atanh}^{-1} \left\{
\frac{ K^{(A)}  \frac{\left(1-\mu^{(A)}\right)^2}{1-2\mu^{(A)}} + K^{(B)} \frac{\left(1-\mu^{(B)}\right)^2}{1-2\mu^{(B)}}-\sqrt{\Delta}}
{2 K^{(A)}K^{(B)} \left[ \frac{\left(1-\mu^{(A)}\right)^2}{1-2\mu^{(A)}} \frac{\left(1-\mu^{(B)}\right)^2}{1-2\mu^{(B)}}-\rho^{2} \right]}
\right\},
\label{Tc2}
\end{equation}
where
\begin{displaymath}
\Delta = \left[ K^{(A)}  \frac{\left(1-\mu^{(A)}\right)^2}{1-2\mu^{(A)}} - K^{(B)} \frac{\left(1-\mu^{(B)}\right)^2}{1-2\mu^{(B)}} \right]^2
+4 K^{(A)}K^{(B)} \rho^{2},
\end{displaymath}
and $\rho=1$ according to Eq.\ (\ref{expval}).
Thus, for the Ising model on MNs with layers in the form of SF networks with the distributions of the degrees of nodes obeying power scaling laws with the
exponents $\gamma^{(A)}=(1-\mu^{(A)})^{-1}>3$, $\gamma^{(B)}=(1-\mu^{(B)})^{-1}>3$, 
the replica approach predicts a finite value of the critical temperature for the FM transition in the thermodynamic limit.
In contrast, if $\mu^{(A)}>1/2$ ($\gamma^{(A)}<3$) or $\mu^{(B)}>1/2$ ($\gamma^{(B)}<3$) the corresponding sums in Eq.\ (\ref{sump2}) diverge, 
and the critical temperature also diverges, so that in the thermodynamic limit the system remains in the FM phase at any temperature, as
predicted using the MF approximation, too (Sec.\ 3.4.2).

\subsubsection{Mutually correlated scale-free layers}

\begin{figure}
\centerline{\includegraphics[width=0.5\textwidth]{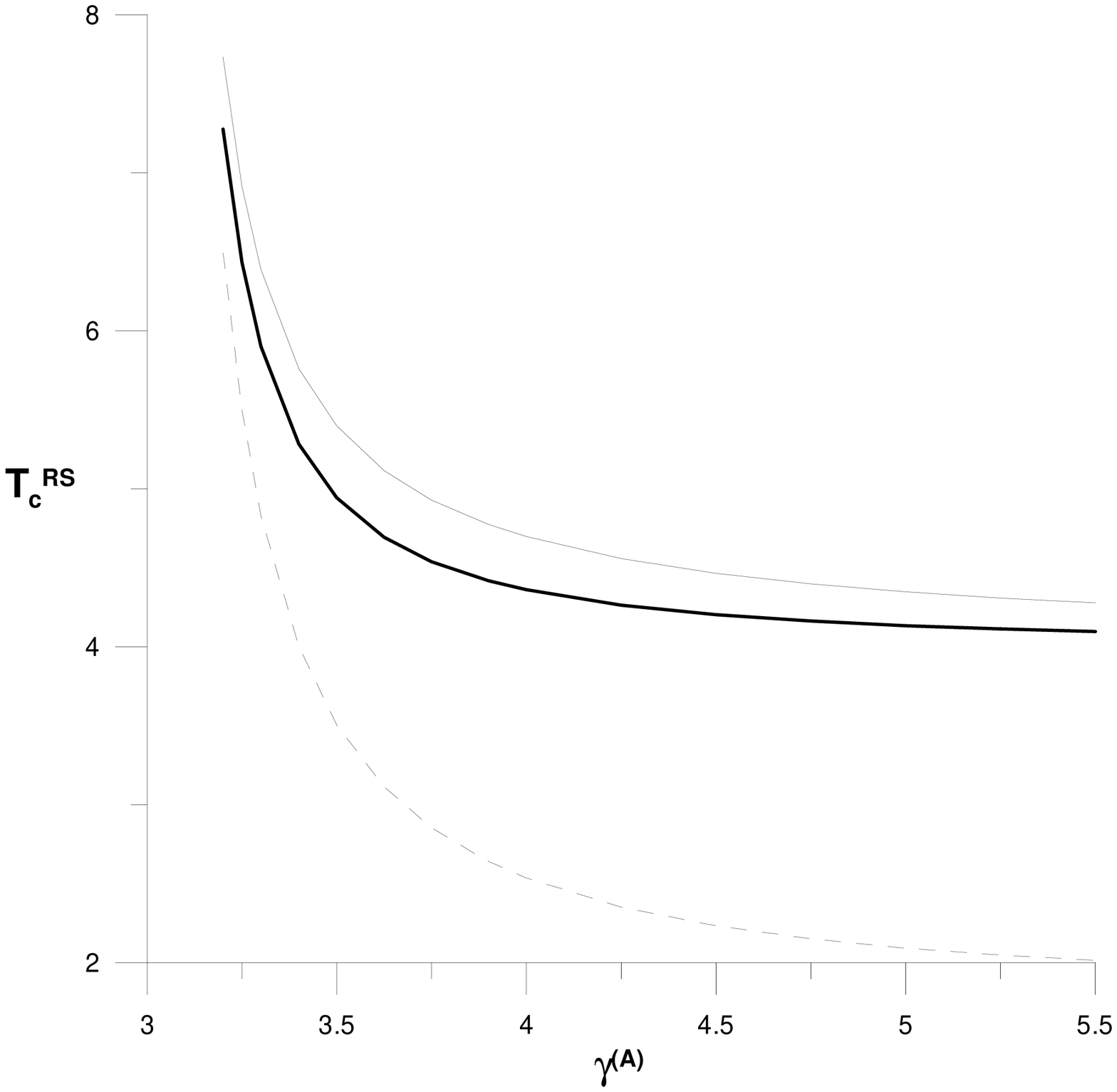}}
\caption{Critical temperature $T_{c}^{RS}$ vs.\ $\gamma^{(A)}$ from the RS solution for the MN with SF layers with $K^{(A)}=K^{(B)}=2$, 
$\gamma^{(B)}=5.5$ and independent layers (black solid line), layers with maximum (gray solid line)  and minimum (gray dashed line) 
correlation between sequences of weights.}
\end{figure}

So far the focus has been on the case of the Ising model on MNs with independently generated layers. 
However, if the layers are generated from the static model it is easy to introduce correlations between the degrees of nodes in
different layers by associating appropriately the two sets of weights with the nodes; thus, here the effect of such correlations on the 
FM transition is briefly discussed. 

In the case of two SF layers maximum correlation between the two sequences of weights, and
thus between mean degrees of each node $\langle k_{i}^{(A)}\rangle$, $\langle k_{i}^{(B)}\rangle$ 
in the two layers in the ensemble of generated layers, is achieved by fixing the numbering of nodes $i=1,2\ldots N$, the same in both layers, and
associating the weights $v_{i}^{(A)}=i^{-\mu^{(A)}}/\zeta_{N}(\mu^{(A)})$, $v_{i}^{(B)}=i^{-\mu^{(B)}}/\zeta_{N}(\mu^{(B)})$ with
the node with index $i$. As a result, the nodes which have high degree within one layer have also, on average, high degree in the other layer and
vice versa. Then for $\mu^{(A)}<1/2$ ($\gamma^{(A)}>3$), $\mu^{(B)}<1/2$ ($\gamma^{(B)}>3$) 
\begin{equation}
\rho=N\sum_{i=1}^{N} v_{i}^{(A)}v_{i}^{(B)} \approx \frac{\left(1-\mu^{(A)}\right) \left(1-\mu^{(B)}\right)}{1-\left( \mu^{(A)}+ \mu^{(B)} \right)}
\label{correl}
\end{equation}
in Eq.\ (\ref{mAmB}). Since all MNs generated in this way are equivalent up to the permutation of the indices of nodes,
no further averaging as in Eq.\ (\ref{expval}) is necessary, and the critical temperature for the MF transition can be obtained by
inserting Eq.\ (\ref{correl}) into Eq.\ (\ref{mAmB}) and equating the determinant to zero; the resulting $T_{c}^{RS}$ is again given by
Eq.\ (\ref{Tc2}) with $\rho$ given by Eq.\ (\ref{correl}).
The critical temperature in this case is higher than that in the case of independently generated SF layers (Fig.\ 2).

In contrast, minimum correlation between the two sequences of weights is obtained if, for fixed numbering of nodes $i=1,2\ldots N$, the same in both layers,
the weights in the two layers are assumed as $v_{i}=i^{-\mu^{(A)}}/\zeta_{N}(\mu^{(A)})$,
$v_{N-i}^{(B)}=i^{-\mu^{(B)}}/\zeta_{N}(\mu^{(B)})$. As a result, 
 the nodes which have high degree within one layer have, on average, low degree in the other layer and
$N\sum_{i=1}^{N} v_{i}^{(A)}v_{i}^{(B)} \stackrel{N\rightarrow \infty}{\rightarrow} 0$.
After inserting this result  in Eq.\ (\ref{mAmB}) it can be seen that in the thermodynamic limit the Ising model on the MN
is decomposed into two practically non-interacting systems on separate SF networks corresponding to the two layers $G^{(A)}$, $G^{(B)}$,
with their own critical temperatures,
\begin{displaymath}
T_{c}^{(A)}= J {\rm atanh}^{-1} \left\{ \frac{\left( \gamma^{(A)}-1\right) \left( \gamma^{(A)} -3 \right)}{\left( \gamma^{(A)}-2\right)^{2}} \right\},
\end{displaymath}
and $T_{c}^{(B)}$ resulting from a formula with $(A)$ replaced with $(B)$. The critical temperature for the FM transition for the whole system is
$T_{c}^{RS}= \max \left\{ T_{c}^{(A)}, T_{c}^{(B)} \right\}$ 
and is lower than that in the case of independently generated SF layers (Fig.\ 2).

Thus, the critical temperature of the Ising model on a MN with SF layers generated from the static model 
depends on the correlation between the weights of nodes in different layers. However, in most cases $T_{c}^{RS}$ should
be close to that for the Ising model on a MN with independent SF layers, Eq.\ (\ref{Tc2}) with $\rho=1$.

\subsubsection{General heterogeneous layers }

In networks generated from the static model there is $\langle k\rangle =K$, 
$N\sum_{i} v_{i}^{2} =\left( \langle k^{2}\rangle -\langle k\rangle\right)/\langle k\rangle^2$ \cite{Lee04}. Thus the result of Eq.\ (\ref{Tc2}) 
with $\rho=1$ can be written in a more general form,
\begin{equation}
T_{c}^{RS}=J {\rm atanh}^{-1} \left\{
\frac{\frac{\langle k^{(A)2}\rangle}{\langle k^{(A)} \rangle} + \frac{\langle k^{(B)2}\rangle}{\langle k^{(B)} \rangle} -2 -\sqrt{\Delta}}
{2 \left[ \left( \frac{\langle k^{(A)2}\rangle}{\langle k^{(A)} \rangle}-1\right) \left( \frac{\langle k^{(B)2}\rangle}{\langle k^{(B)} \rangle} -1\right)
-\langle k^{(A)} \rangle \langle k^{(B)} \rangle\right]}
\right\}
\label{Tc3}
\end{equation}
where
\begin{displaymath}
\Delta = \left( \frac{\langle k^{(A)2}\rangle}{\langle k^{(A)} \rangle} - \frac{\langle k^{(B)2}\rangle}{\langle k^{(B)} \rangle}\right)^{2}
+4 \langle k^{(A)} \rangle \langle k^{(B)} \rangle,
\end{displaymath}
using the moments of the distributions of the degrees of nodes within each layer. 
It can be expected that Eq.\ (\ref{Tc3}) is valid for any multiplex
network consisting of independently generated, possibly heterogeneous
layers with finite second moments of the distributions of the degrees of nodes. 
This case corresponds to that cosnidered in the framework of the MF theory in Sec.\ 3, where the joint distributions of the degrees of
nodes in the MN were products of distributions within each layer, $p_{k^{(A)},k^{(B)}} = p_{k^{(A)}} p_{k^{(B)}}$.

It should be noted that the necessary condition for the occurrence of the FM transition is that the critical temperature $T_{c}^{RS}$
given by Eq.\ (\ref{Tc3}) is real and positive. It can be easily shown that this requires that
\begin{equation}
\frac{\langle k^{(A)2}\rangle}{\langle k^{(A)} \rangle} + \frac{\langle k^{(B)2}\rangle}{\langle k^{(B)} \rangle} +\sqrt{\Delta} >4,
\end{equation}
which is also a condition for the occurrence of a giant component  in a MN with two independently generated layers 
\cite{Lee14}, in which nodes are connected via edges in any layer (but not necessarily in both layers). Thus, the FM transition can appear
in MNs above the percolation threshold, in analogy with the case of complex networks \cite{Kim05}.

As a particular case of Eq.\ (\ref{Tc3}) let us consider the critical temperature for the Ising model on
a MN consisting of two layers with identical distributions of the
degrees of nodes, i.e., $p_{k^{(A)}}=p_{k^{(B)}}=p_{k}$, and thus $\langle k^{(A)}\rangle =\langle k^{(B)} \rangle =\langle k\rangle$,
$\langle k^{(A)2}\rangle =\langle k^{(B)2} \rangle =\langle k^{2}\rangle$. Then
\begin{equation}
T_{c}^{RS}= J {\rm atanh}^{-1} \left[ \left( \frac{\langle k^{2} \rangle}{\langle k\rangle} +\langle k\rangle -1\right)^{-1} \right]
=-2J\ln^{-1} \left( 1- \frac{2}{\frac{\langle k^{2} \rangle}{\langle k\rangle}+ \langle k\rangle} \right).
\end{equation}
For large $\langle k^{2}\rangle$, $\langle k\rangle$ the critical temperature can be approximated as
\begin{equation}
T_{c}^{RS} \approx J\left( \frac{\langle k^{2} \rangle}{\langle k\rangle}+ \langle k\rangle\right),
\end{equation}
which is the MF result, Eq.\ (\ref{Tcmfident}), as expected for layers with large mean degrees of nodes.

In particular, if the distributions of the degrees of nodes within each layer obey power scaling laws in the form
$p_{k^{(A)}} = \left( \gamma^{(A)}-1\right) \tilde{m}^{\gamma^{(A)}-1} \left( k^{(A)}\right)^{-\gamma^{(A)}}$ for $k^{(A)}>\tilde{m}$, 
$p_{k^{(B)}} = \left( \gamma^{(B)}-1\right) \tilde{m}^{\gamma^{(B)}-1} \left( k^{(B)}\right)^{-\gamma^{(B)}}$ for $k^{(B)}>\tilde{m}$, 
the critical temperature can be obtained by inserting in Eq.\ ({\ref{Tc3})
\begin{equation}
\frac{\langle k^{(A)2}\rangle}{\langle k^{(A)} \rangle}= \tilde{m} \frac{\gamma^{(A)}-2}{\gamma^{(A)}-3}, \;\; 
\langle k^{(A)} \rangle = \tilde{m} \frac{\gamma^{(A)}-1}{\gamma^{(A)}-2},
\label{moments}
\end{equation}
and similar expressions for the moments of $p_{k^{(B)}}$. The result can be compared with Eq.\ (\ref{TcmfSF}) obtained in the MF approximation
(see Sec.\ 4.6).

\subsection{Comparison with the mean field theory and Monte Carlo simulations}

In this section the critical temperature for the FM transition in the Ising model on MNs with independent SF layers obtained from the RS solution is
compared to the corresponding MF critical temperature, Eq.\ (\ref{TcmfSF}), and to that obtained from MC simulations of the Ising model
on MNs with independent SF layers generated from the Configuration Model (Sec.\ 3.5). Since the layers are independent the appropriate
formula for $T_{c}^{RS}$ is given by Eq.\ (\ref{Tc3}) with the moments of the distributions of the degrees of nodes given by 
Eq.\ (\ref{moments}). In Fig.\ 1(c) $T_{c}^{RS}$ is shown 
for the Ising model on MNs with SF layers with $J=1$ and with fixed $\gamma^{(A)}>3$, $\gamma^{(B)}>3$ and different $\tilde{m}$
and compared with $T_{c}^{MF}$ and $T_{c}^{MC}$. Though Eq.\ (\ref{Tc3}) is only a heuristic generalization 
of Eq.\ (\ref{Tc2}) to the case of arbitrary heterogeneous layers 
$T_{c}^{RS}$ shows better quantitative agreement with $T_{c}^{MC}$ than the MF critical temperature $T_{c}^{MF}$. 
The discrepancy between the analytic result from the RS solution and the result of MC simulations can be probably again attributed
mainly to the fact that in the calculations leading to Eq.\ (\ref{Tc2}) the distributions $p_{k^{(A)}}$, $p_{k^{(B)}}$ were assumed
continuous (cf.\ Eq.\ (\ref{sump2})). Besides, $T_{c}^{RS}$ shows the same linear dependence on $\tilde{m}$ as $T_{c}^{MF}$,
predicted by Eq.\ (\ref{TcmfSF}), and as $T_{c}^{MC}$.

In Fig.\ 1(d) $T_{c}^{RS}$ is shown for the Ising model on MNs with SF layers with $J=1$, fixed $\gamma^{(B)}=5.5$ and
high $\tilde{m}=20$ and with different $\gamma^{(A)}>3$ and compared with $T_{c}^{MF}$ and $T_{c}^{MC}$. 
It can be seen that  $T_{c}^{RS}$ diverges as $\gamma^{(A)}\rightarrow 3$, as expected, and the MF critical temperature 
approaches that obtained from the RS solution. In the whole range of $\gamma^{(A)}$ the critical temperature $T_{c}^{RS}$
shows better quantitative agreement with $T_{c}^{MC}$ than the MF value $T_{c}^{MF}$. 

The qualitative, and to some extent even quantitative agreement between $T_{c}^{MF}$ and the more rigorously evaluated
$T_{c}^{RS}$ confirms the validity of the MF approximation in the investigation of the Ising model on MNs with high enough
density of connections.

\section{Critical behavior of the magnetization}

Below the transition point from the paramagnetic to the FM phase the magnetization is expected to increase from zero as
$\varepsilon^{\beta}$, where $\varepsilon =\left( T_{c}-T\right)/T_{c}$.
Besides, for the Ising model on SF networks with diverging second moment of the distribution of the degrees of nodes
the weighted magnetization is expected to decrease as $T^{-\alpha}$ as $T\rightarrow \infty$.
In Ref.\ \cite{Leone02,Kim05} using methods from the SG theory it was shown that in the case of the Ising model on 
SF networks the scaling exponents $\alpha$, $\beta$ can be non-universal and
depend on the parameters of the distribution of the degrees of nodes. In this section
these exponents are evaluated for the Ising model on MNs with independent SF layers using the RS solution obtained from Eq.\ (\ref{system}).

In order to obtain scaling for the weighted magnetization the right-hand sides of the equations in Eq.\ (\ref{system}) should be 
expanded with respect to the powers of $m^{(A)}$, $m^{(B)}$. 
Unfortunately, in this case it is not possible simply to expand the $\tanh \left( \cdot\right)$ function in the Taylor series
due to the occurrence of terms like $N^{-1}\sum_{i=1}^{N} v_{i}^{(A)3}$, etc., which diverge even if the second moments of
the distributions of the weights associated with each layer are finite. Nevertheless, as shown in the Appendix,
the sums over the indices of nodes on the right-hand sides of Eq.\ (\ref{system}) can be represented in a form of a converging series expansion
with respect to $m^{(A)}$, $m^{(B)}$. For this purpose, let us note that in the case of independent SF layers these sums can be
replaced by their expected values, similarly as in Sec.\ 4.5.2, Eq.\ (\ref{expval}), and then approximated by an integral, e.g., 
\begin{eqnarray}
&& \sum_{i=1}^{N} v_{i}^{(A)} 
\tanh\left[  N  \left( K^{(A)}{\bf T}_{1} v_{i}^{(A)} m^{(A)} + K^{(B)} {\bf T}_{1} v_{i}^{(B)} m^{(B)}\right) \right] 
\approx \nonumber\\
&& \sum_{i=1}^{N}  N^{-2} \sum_{k=1}^{N}\sum_{l=1}^{N} v_{k}
\tanh \left[  N  \left( K^{(A)}{\bf T}_{1} v_{k} m^{(A)} + K^{(B)} {\bf T}_{1} v_{l} m^{(B)}\right) \right]
\approx \nonumber\\
&& \frac{1-\mu^{(A)}}{N^{2}} \int_{1}^{N}\int_{1}^{N} dy_{k} dy_{l} \left( \frac{N}{y_{k}}\right)^{\mu^{(A)}}
\tanh \left[  \left( \frac{N}{y_{k}}\right)^{\mu^{(A)}} M^{(A)} +  \left( \frac{N}{y_{l}}\right)^{\mu^{(B)}} M^{(B)}  \right],
\nonumber
\end{eqnarray}
where $M^{(A)}=\left( 1-\mu^{(A)}\right) K^{(A)}{\bf T}_{1} m^{(A)}$,  $M^{(B)}=\left( 1-\mu^{(B)}\right) K^{(B)}{\bf T}_{1} m^{(B)}$.
In the limit $N\rightarrow \infty$ and after replacing the variables $u_{1}=M^{(A)} \left(N/y_{k}\right)^{\mu^{(A)}}$,
$u_{2}=M^{(B)} \left(N/y_{l}\right)^{\mu^{(B)}}$ Eq.\ (\ref{system}) becomes
\begin{eqnarray}
\frac{m^{(A)}}{1-\mu^{(A)}}& =& \left( \gamma^{(A)}-1\right) \left( \gamma^{(B)}-1\right) 
\left( M^{(A)}\right)^{\gamma^{(A)}-2} \left( M^{(B)}\right)^{\gamma^{(B)}-1}
\times \nonumber\\
&&\int_{M^{(A)}}^{\infty} \int_{M^{(B)}}^{\infty} 
\frac{u_{1}\tanh \left( u_{1}+u_{2} \right) }{u_{1}^{\gamma^{(A)}} u_{2}^{\gamma^{(B)}}} du_{2} du_{1},
\label{mAfm2}
\end{eqnarray}
and analogous equation for $M^{(B)}$. The two-dimesional integral in Eq.\ (\ref{mAfm2}) can be evaluated using Eq.\ (\ref{s}) with 
$F\left( x_{1},x_{2}\right) = x_{1}\tanh \left( x_{1}+x_{2}\right)$ in the Appendix, and the result is given by Eq.\ (\ref{Sfinal}).
Inserting this result in Eq.\ (\ref{mAfm2}) and retaining only important nonlinear terms of maximum order $M^{3}$ (the terms of order $M^{2}$
are absent since $f_{3,0}=f_{2,1}=f_{1,2}=0$ in Eq.\ (\ref{Sfinal}), see Eq.\ (\ref{Taylorcoeff}))
the following system of nonlinear equations for the order parameters is obtained, valid for non-integer $\gamma^{(A)}>2$, $\gamma^{(B)}>2$,
\begin{eqnarray}
&& \left[\frac{1}{K^{(A)}{\bf T}_{1}}
- \frac{\left( \gamma^{(A)}-2\right)^{2}}{\left( \gamma^{(A)}-1\right)\left( \gamma^{(A)}-3\right)} \right]M^{(A)}
- \frac{\left( \gamma^{(A)}-2\right)\left( \gamma^{(B)}-1\right)}{\left( \gamma^{(A)}-1\right)\left( \gamma^{(B)}-2\right)} M^{(B)} 
\nonumber  \\
&& = \frac{\left( \gamma^{(A)}-2\right)^{2}}{\gamma^{(A)}-1} I_{1}\left(\gamma^{(A)},0\right) 
\left( M^{(A)}\right)^{\gamma^{(A)}-2}  
\nonumber\\
&&+\frac{\left( \gamma^{(A)}-2\right) \left( \gamma^{(B)}-1\right)}{\gamma^{(A)}-1}
I_{2}\left(\gamma^{(B)},1\right) \left( M^{(B)}\right)^{\gamma^{(B)}-1}
\nonumber\\
&&- \frac{1}{3}\frac{\left( \gamma^{(A)}-2\right)^{2}}{\left( \gamma^{(A)}-1\right)\left( \gamma^{(A)}-5\right)}M^{(A)3}
- \frac{\left( \gamma^{(A)}-2\right)^{2}\left( \gamma^{(B)}-1\right)}
{\left( \gamma^{(A)}-1\right) \left( \gamma^{(A)}-4\right)\left( \gamma^{(B)}-2\right)} M^{(A)2}M^{(B)}
\nonumber\\
&& -\frac{\left( \gamma^{(A)}-2\right)^{2}\left( \gamma^{(B)}-1\right)}
{\left( \gamma^{(A)}-1\right) \left( \gamma^{(A)}-3\right)\left( \gamma^{(B)}-3\right)} M^{(A)}M^{(B)2}
- \frac{ \left( \gamma^{(A)}-2\right)\left( \gamma^{(B)}-1\right)}{\left( \gamma^{(A)}-1\right)\left( \gamma^{(B)}-4\right)} M^{(B)3}
\nonumber\\
&&
\label{mAmBfm2}
\end{eqnarray}
and a complementary equation which can be obtained from Eq.\ (\ref{mAmBfm2}) by replacing $(A)$ with $(B)$ 
and vice versa, and
\begin{eqnarray}
I_{1}\left( \lambda,0\right)&=& \left\{
\begin{array}{ccc}
 \int_{0}^{\infty} x^{1-\lambda}\tanh x dx & {\rm for}& 1<\gamma^{(A)} <3 \\
 \int_{0}^{\infty} x^{1-\lambda} \left(\tanh x-x \right) dx & {\rm for}& 3<\gamma^{(A)} <5 
\end{array}
\right.
\nonumber
\end{eqnarray}
\begin{eqnarray}
I_{2}\left( \lambda,1\right)&=& \left\{
\begin{array}{ccc}
 \int_{0}^{\infty} x^{-\lambda} \left(\tanh x-x \right) dx & {\rm for}& 2<\gamma^{(A)} <4 \\
 \int_{0}^{\infty} x^{-\lambda} \left(\tanh x-x +x^3 \right) dx & {\rm for}& 4<\gamma^{(A)} <5. 
\end{array}
\right.
\nonumber
\end{eqnarray}
Let us note that the left-hand (linear) part of the above-mentioned system of equations is identical with that of
Eq.\ (\ref{mAmB}) in the case of independent layers, with $\rho = \sum_{i=1}^{N}v_{i}^{(A)}v_{i}^{(B)} \rightarrow 1$.
The remaining nonlinear terms are never dominant in Eq.\ (\ref{Sfinal}) for $\gamma^{(A)}>2$, $\gamma^{(B)}>2$, thus can be omitted.
For integer $\gamma^{(A)}$ or $\gamma^{(B)}$ terms with logarithmic corrections of scaling occur in Eq.\ (\ref{mAmBfm2}); this case is not
discussed here for the sake of brevity.

For $\gamma^{(A)}>3$, $\gamma^{(B)}>3$ the critical temperature $T_{c}^{RS}$ is finite and Eq.\ (\ref{mAmBfm2}) can be used to
find the critical behavior of the weighted magnetizations for $\varepsilon \rightarrow 0$.
For $3< \gamma^{(A)}<5$ and $\gamma^{(B)}>\gamma^{(A)}$, assuming that 
the small quantities $M^{(A)}$  and $M^{(B)}$ are of the same odrer of magnitude,
the dominat terms in Eq.\ (\ref{mAmBfm2}) are the nonlinear term with $\left( M^{(A)} \right)^{\gamma^{(A)}-2}$
and the linear terms with $M^{(A)}$, $M^{(B)}$, and the remaining nonlinear terms can be neglected.
It is then possible to evaluate from this equation $M^{(B)} =C_{1} M^{(A)} \left(1+C_{2} \left( M^{(A)}\right)^{\gamma^{(A)}-3}\right)$,
where the constants $C_{1}$, $C_{2}$ are to leading order independent of $\varepsilon$. After inserting this formula in the equation
complementary to Eq.\ (\ref{mAmBfm2}) and expanding $\left( M^{(B)} \right)^{\gamma^{(B)}-2}$ in powers of the small term 
$\left( M^{(A)}\right)^{\gamma^{(A)}-3}$ an equation in a general form $C(\varepsilon ) + C_{3} \left( M^{(A)}\right)^{\gamma^{(A)}-3}=
O\left( \left( M^{(A)}\right)^{\delta}\right)$ is obtained, where $C(\varepsilon )$ denotes the determinant of the linear part of the 
system of equations in Eq. (\ref{mAmBfm2}), $C_{3}$ is a constant to leading order independent of $\varepsilon$,
and the terms on the right-hand side are of order $\delta > \gamma^{(A)}-3$ and thus can be neglected. Since $C(0)=0$ in the
first approximation there is $C(\varepsilon )\propto \varepsilon$ and the scaling behavior of the order parameters is obtained as
$m^{(A)}\propto M^{(A)}\propto \varepsilon ^{\frac{1}{\gamma^{(A)}-3}}$, 
$m^{(B)}\propto m^{(A)}\propto \varepsilon ^{\frac{1}{\gamma^{(A)}-3}}$. Similarly, 
for $3< \gamma^{(B)}<5$ and $\gamma^{(A)}>\gamma^{(B)}$ the scaling behavior is 
$m^{(A)}\propto \varepsilon ^{\frac{1}{\gamma^{(B)}-3}}$, $m^{(B)}\propto \varepsilon ^{\frac{1}{\gamma^{(B)}-3}}$. Hence,
if $3< \gamma^{(A)} <5$ or $3< \gamma^{(B)} <5$ the expected scaling behavior for the magnetization in the vicinity of $T_{c}^{RS}$ is
$m^{(A,B)} \propto \varepsilon ^{\frac{1}{\gamma_{min}-3}}$, where $\gamma_{min}=\min \left\{ \gamma^{(A)}, \gamma^{(B)} \right\}$,
i.e., it is determined by the more heterogeneous layer.

For $\gamma^{(A)}>5$, $\gamma^{(B)}>5$ the lowest-order nonlinear terms in Eq.\ (\ref{mAmBfm2}) are of order $O\left( M^3\right)$ in both
$M^{(A)}$, $M^{(B)}$. It is then not easy to reduce this system of equations to one equation for $M^{(A)}$ or $M^{(B)}$; nevertheless, 
due to the overall form of the nonlinearity it can be expected that the magnetizations should obey the MF scaling relation
$m^{(A,B)}\propto \varepsilon ^{1/2}$.

For $2 < \gamma^{(A)} <3$ or $2 <\gamma^{(B)} <3$ the critical temperature diverges and Eq.\ (\ref{mAmBfm2}) can be used to find the
critical behavior of the weighted magnetization for $T\rightarrow \infty$ and thus for ${\bf T}_{1} =\tanh \beta \approx T^{-1}$. 
If $2< \gamma^{(A)} <3$ and $\gamma^{(B)}\gg 3$ it is reasonable to assume that $M^{(B)} \le M^{(A)}$. Then the dominant terms in 
Eq.\ (\ref{mAmBfm2}) are those with $TM^{(A)}$ and $\left( M^{(A)}\right) ^{\gamma^{(A)}-2}$ and the remaining terms can be neglected.
This yields $M^{(A)} \propto T^{-\frac{1}{3-\gamma^{(A)}}}$ and $m^{(A)} \propto T^{-\frac{\gamma^{(A)}-2}{3-\gamma^{(A)}}}$. In
the equation complementary to Eq.\ (\ref{mAmBfm2}) the dominant terms are those with $M^{(A)}$ and $TM^{(B)}$ which yields
$M^{(B)} \propto T^{-1}M^{(A)} \propto  T^{-\frac{4-\gamma^{(A)}}{3-\gamma^{(A)}}}$ and 
$m^{(B)} \propto T^{-\frac{1}{3-\gamma^{(A)}}}$, i.e., $M^{(B)}$, $m^{(B)}$ tend to zero with $T$ faster than
$M^{(A)}$, $m^{(A)}$. If $2< \gamma^{(A)} \ll 3$ and $2< \gamma^{(B)} \ll 3$ the dominat terms in Eq.\ (\ref{mAmBfm2}) 
and thus the scaling for $M^{(A)}$, $m^{(A)}$ are as above,
and in the complementary equation the terms with $TM^{(B)}$, $\left( M^{(B)}\right) ^{\gamma^{(B)}-2}$ are dominant, thus the scaling 
$M^{(B)} \propto T^{-\frac{1}{3-\gamma^{(B)}}}$ and $m^{(B)} \propto T^{-\frac{\gamma^{(B)}-2}{3-\gamma^{(B)}}}$ is expected.
However, it should be noted that, e.g., for fixed $2< \gamma^{(A)} <3$ the predictions for the scaling exponents for 
$M^{(B)}$, $m^{(B)}$ are inconsistent as $\gamma^{(B)} \rightarrow 3^{\pm}$. Hence, in this region a sort of crossover scaling behavior is
expected. This is since $M^{(A)}$ and $M^{(B)}$ can scale in a different way and thus in the equation complementary to Eq.\ (\ref{mAmBfm2})
the terms $M^{(A)}$ and $\left( M^{(B)}\right) ^{\gamma^{(B)}-2}$ can be comparable for $\gamma^{(B)} \rightarrow 3^{\pm}$.
Similar argument shows that for $2 \ll \gamma^{(A)} <3$, $2 \ll \gamma^{(B)} <3$ corrections to the scaling
$M^{(A)} \propto T^{-\frac{1}{3-\gamma^{(A)}}}$, $M^{(B)} \propto T^{-\frac{1}{3-\gamma^{(B)}}}$ occur.
To summarize, for $2 < \gamma^{(A)} <3$ or $2 <\gamma^{(B)} <3$ and for $T\rightarrow \infty$
the scaling for the dominant (more slowly decreasing to zero) component of the weighted magnetization 
$M_{max}=\max \left\{ M^{(A)}, M^{(B)} \right\}$ and $m_{max}= \max \left\{ m^{(A)}, m^{(B)} \right\}$ 
is $M_{max}\propto T^{-\frac{1}{3-\gamma_{min}}}$, $m_{max}\propto T^{-\frac{\gamma_{min}-2}{3-\gamma_{min}}}$
with possible corrections discussed above, i.e., it is again determined by the more heterogeneous layer.

The predicted scaling behavior for the weighted magnetization for the Ising model with independent SF layers is summarized in Table I.
It is interesting to note that the same scaling behavior can be expected by considering the Ising model on a corresponding super-network 
for which $p_{k}\propto k^{-\gamma_{min}}$ up to leading term \cite{Kim05}. For completeness, it should be mentioned that 
severe mathematical difficulties were faced while trying to determine the critical properties of the 
Ising model on mutually (in particular, maximally) correlated SF layers. Hence, this case is not discussed in this paper.

\newpage

\begin{center}
\begin{tabular}{c|ccc}
\hline 
 & $2 < \gamma_{min} < 3$ & $ 3 < \gamma_{min} <5$ & $ \gamma_{min}>5$ \\
\hline
$m_{max}$ & $T^{-\frac{\gamma_{min}-2}{3-\gamma_{min}}}\star$   & $\varepsilon^{\frac{1}{\gamma_{min}-3}}$  & $\varepsilon^{1/2}$ \\
$M_{max}$ & $T^{-\frac{1}{3-\gamma_{min}}}\star$ & $\varepsilon^{\frac{1}{\gamma_{min}-3}}$ & $\varepsilon^{1/2}$
\end{tabular}
\end{center}
{\footnotesize $^{\star}$ corrections expected for $\gamma^{(A,B)}\rightarrow 3^{-}$}

{\footnotesize Table I. Scaling behavior for the dominant part of the magnetization
$m_{max}= \max \left\{ m^{(A)}, m^{(B)} \right\}$, $M_{max}=\max \left\{ M^{(A)}, M^{(B)} \right\}$.
}

\section{Summary and conclusions}

In this paper a simple version of the FM Ising model was investigated on, possibly heterogeneous, MNs with separately generated layers which can have
different structural properties and distributions of the degrees of nodes. Critical temperatures for the FM transition were evaluated
from the MF approximation and using the replica method, from the RS solution, in particular in the case of random ER and SF layers, 
and compared with results of MC simulations. In the case of independently generated layers
the two above-mentioned analytic methods yield qulalitatively similar results for
MNs with high density of connections, and the critical temperature obtained from the RS solution shows better quantitative
agreement with numerical results. Using the replica approach it was also shown that the critical temperature depends sensitively
on the correlations between the degrees of nodes in different layers: it is increased by positive and decreased by negative correlated multiplexity.
This result is analogous to the observations of the lowering of the point of the mutual percolation and  the related increase of the robustness 
against random failures \cite{Lee12,Min14} and mitigating the cascading failures \cite{Tan13}
in systems on MNs in the case of positive correlations between degrees of nodes in different layers. Investigation of the Ising model on MNs
both in the framework of MF approximation and the replica method requires using different "partial" order parameters
or magnetizations, respectively, associated with different layers of the MN, thus the study of the properties of the
phase transition is a more difficult task than in the case of the Ising model on complex (possibly heterogeneous) networks.
Nevertheless, critical exponents for the dominant component of the weighted magnetization were found for the Ising model on 
a MN with independent SF layers and it was shown that for strongly heterogeneous layers they are determined by the properties of 
the distribution of the degrees of nodes of the most heterogeneous layer.

The study of such a basic and simple model as the variant of the Ising model on MNs 
considered in this paper does not lead to such qualitatively new results as, e.g., the occurrence of discontinuous phase transitions
in the mutual percolation problem \cite{Buldyrev10,Baxter12}, threshold cascades \cite{Lee14a} and  the Ashkin-Teller model \cite{Jang15}.
Nevertheless, it reveals various quantitative effects of the 
multiplex structure of the network of interactions on the properties of continuous phase transitions, thus emphasising
inaccuracy of their description based on the super-network only, with the underlying structure completely neglected. For example, already
in the framework of the simple heterogeneous MF theory, even if the FM exchange interactions within all layers are equal, noticeable 
differences appear between the critical temperatures
for the FM transition obtained for the Ising model on a MN with strongly heterogeneous layers and on a super-network with the correlations between
the degrees of nodes induced by the separate generation of layers neglected. On the other hand, both above-mentioned approaches
predict the same critical behavior of the weighted magnetization.

The heterogeneous MF theory presented in this paper can be easily generalized to the case of the Ising model on partly overlapping MNs.
Using a similar approach it should be also possible to investigate phase transitions in other systems on MNs, in particular 
with heterogeneous layers. The replica method can be applied to study the possible SG transition in the Ising model on MNs with
both FM and antiferromagnetic interactions.
Another challenging problem is determination of the critical exponents for the two subsets of 
the order parameters associated with the two heterogeneous layers of the MN at the FM and SG transition points.
In the above-mentioned cases results of this paper provide a starting point for the future research.

\section*{Appendix}

In this Appendix a general expansion formula is derived for integrals of a form
\begin{equation}
S\left( y_{A},y_{B}\right) = \left( \lambda_{A}-1\right) \left( \lambda_{B}-1\right) y_{A}^{\lambda_{A}-2} y_{B}^{\lambda_{B}-1} 
\int_{y_{A}}^{\infty} \int_{y_{B}}^{\infty} \frac{F\left( x_{1},x_{2}\right) }{x_{1}^{\lambda_{A}} x_{2}^{\lambda_{B}}} dx_{2} dx_{1},
\label{s}
\end{equation}
using a method which is a generalization of that from Ref.\ \cite{Kim05} to the case of two-dimensional integrals. It is assumed that
$F\left( x_{1},x_{2}\right)$ is a differentiable function which diverges slower than $x_{1}^{\lambda_{1}-1}$ and $x_{2}^{\lambda_{2}-1}$.

Let us assume that $\lambda_{A}$, $\lambda_{B}$ are not integer numbers and, for some $m_{1}$, $m_{2}$ there is $m_{1} < \lambda_{A} < m_{1}+1$,
$m_{2} < \lambda_{B} < m_{2} +1$ (for integer $\gamma^{(A)}$ or $\gamma^{(B)}$ logarithmic corrections to the formulae
derived below are expected \cite{Kim05}, which are not discussed here for the sake of brevity). 
Expansion of the function $F\left( x_{1},x_{2}\right)$ in the Taylor series is
\begin{eqnarray}
F\left( x_{1},x_{2}\right) &= & \sum_{n_{1},n_{2}=0}^{\infty} f_{n_{1},n_{2}} x_{1}^{n_{1}}x_{2}^{n_{2}}=  \nonumber \\
&=& \sum_{n_{1}=0}^{m_{1}-1} \sum_{n_{2}=0}^{m_{2}-1} f_{n_{1},n_{2}} x_{1}^{n_{1}}x_{2}^{n_{2}} \nonumber \\
&+&  \sum_{n_{1}=m_{1}}^{\infty} \sum_{n_{2}=0}^{m_{2}-1} f_{n_{1},n_{2}} x_{1}^{n_{1}}x_{2}^{n_{2}}
+ \sum_{n_{1}=0}^{m_{1}-1} \sum_{n_{2}=m_{2}}^{\infty} f_{n_{1},n_{2}} x_{1}^{n_{1}}x_{2}^{n_{2}} \nonumber \\
&+&  \sum_{n_{1}=m_{1}}^{\infty} \sum_{n_{2}=m_{2}}^{\infty} f_{n_{1},n_{2}} x_{1}^{n_{1}}x_{2}^{n_{2}},
\label{Taylorexp}
\end{eqnarray}
where the expansion coefficients are
\begin{equation}
f_{n_{1},n_{2}} = \frac{1}{n_{1}! n_{2}!} 
\left. \frac{\partial ^{n_{1}+n_{2}} F}{\partial x_{1}^{n_{1}} \partial x_{2}^{n_{2}}} \left( x_{1},x_{2}\right) \right|_{(0,0)}.
\label{Taylorcoeff}
\end{equation}
The first sum in Eq.\ (\ref{Taylorexp}) can be integrated term by term which yields
\begin{eqnarray}
&& \sum_{n_{1}=0}^{m_{1}-1} \sum_{n_{2}=0}^{m_{2}-1} f_{n_{1},n_{2}} 
\int_{y_{A}}^{\infty} x_{1}^{n_{1}-\lambda_{A}} dx_{1} \int_{y_{B}}^{\infty} x_{2}^{n_{2}-\lambda_{B}} dx_{2} \nonumber \\
&& = \sum_{n_{1}=0}^{m_{1}-1} \sum_{n_{2}=0}^{m_{2}-1} f_{n_{1},n_{2}}
\frac{(-1)}{n_{1}-\lambda_{A}+1} \frac{(-1)}{n_{2}-\lambda_{B}+1} y_{A}^{n_{1}-\lambda_{A}+1} y_{B}^{n_{2}-\lambda_{B}+1}.
\label{c1}
\end{eqnarray}

Concerning the remaining terms it should be noted that in the converging Taylor series, Eq.\ (\ref{Taylorexp}), the order of summation and
integration can be exchanged. Then, e.g., from the integration of the second sum, after evaluating the integral
$\int_{y_{B}}^ {\infty} x_{2}^{n_{2}-\lambda_{B}} dx_{2}= - \left( n_{2}-\lambda_{B} +1 \right)^{-1} y_{B}^{n_{2}-\lambda_{B} +1}$
for $0 \le n_{2} \le m_{2}-1$ and dividing in two parts and evaluating the integral
$\int_{y_{A}}^ {\infty} x_{1}^{n_{1}-\lambda_{A}} dx_{1}=\left( \int_{0}^ {\infty} - \int_{0}^ {y_{A}} \right) x_{1}^{n_{1}-\lambda_{A}} dx_{1}=
\int_{0}^ {\infty}x_{1}^{n_{1}-\lambda_{A}} dx_{1} - \left( n_{1}-\lambda_{A} +1 \right)^{-1} y_{A}^{n_{1}-\lambda_{A} +1}$
for $n_{1} \ge m_{1}$ it is obtained that
\begin{eqnarray}
&& \sum_{n_{1}=m_{1}}^{\infty} \sum_{n_{2}=0}^{m_{2}-1} f_{n_{1},n_{2}} 
\int_{y_{A}}^ {\infty} x_{1}^{n_{1}-\lambda_{A}} dx_{1} \int_{y_{B}}^ {\infty} x_{2}^{n_{2}-\lambda_{B}} dx_{2} = \nonumber \\
&& - \sum_{n_{2}=0}^{m_{2}-1} I_{1}\left( \lambda_{A},n_{2}\right) 
\frac{y_{B}^{ n_{2}-\lambda_{B} +1}}{ n_{2}-\lambda_{B} +1} 
+ \sum_{n_{2}=0}^{m_{2}-1} \sum_{n_{1}=m_{1}}^{\infty} f_{n_{1},n_{2}} 
\frac{y_{A}^{ n_{1}-\lambda_{A} +1}}{ n_{1}-\lambda_{A} +1} \frac{y_{B}^{ n_{2}-\lambda_{B} +1}}{ n_{2}-\lambda_{B} +1}, \nonumber\\
&& \label{c2}
\end{eqnarray}
where
\begin{equation}
 I_{1}\left( \lambda_{A},n_{2}\right) = \sum_{n_{1}=m_{1}} ^{\infty}f_{n_{1},n_{2}} \int_{0}^{\infty} x_{1}^{n_{1}-\lambda_{A}} dx_{1}.
\label{I1}
\end{equation}
Since $m_{1}< \lambda_{A}<m_{1}+1$ the integrals in Eq.\ (\ref{I1}) are not singular and the whole series converges. Taking into account that, from
Eq.\ (\ref{Taylorcoeff}),
\begin{eqnarray}
\sum_{n_{1}=0}^{\infty} f_{n_{1},n_{2}}x_{1}^{n_{1}}&=& \frac{1}{n_{2}!} \sum_{n_{1}=0}^{\infty} \frac{1}{n_{1}!} 
\frac{\partial^{n_{1}}}{\partial x_{1}^{n_{1}}} \left.  \left[ \frac{\partial^{n_{2}}F}{\partial x_{2}^{n_{2}}} \left( x_{1},x_{2} \right)\right] \right|_{(0,0)}
x_{1}^{n_{1}} = \frac{1}{n_{2}!} \left. \frac{\partial^{n_{2}}F}{\partial x_{2}^{n_{2}}} \left( x_{1},x_{2} \right) \right|_{x_{2}=0}, \nonumber\\
&&
\end{eqnarray}
Eq.\ (\ref{I1}) can be rewritten as
\begin{eqnarray}
 I_{1}\left( \lambda_{A},n_{2}\right) &= & \int_{0}^{\infty} x_{1}^{-\lambda_{A}}
\left( \sum_{n_{1}=0}^{\infty} f_{n_{1},n_{2}}x_{1}^{n_{1}} - \sum_{n_{1}=0}^{m_{1}-1} f_{n_{1},n_{2}}x_{1}^{n_{1}} \right) dx_{1} 
\nonumber\\
&=& \int_{0}^{\infty} x_{1}^{-\lambda_{A}}
\left(  \frac{1}{n_{2}!} \left. \frac{\partial^{n_{2}}F}{\partial x_{2}^{n_{2}}} \left( x_{1},x_{2} \right) \right|_{x_{2}=0}
- \sum_{n_{1}=0}^{m_{1}-1} f_{n_{1},n_{2}}x_{1}^{n_{1}} \right) dx_{1}. \nonumber\\
&& \label{I11}
\end{eqnarray}
The third and fourth sum in Eq.\ (\ref{Taylorexp}) can be integrated in a similar way, and finally from Eq.\ (\ref{s}) it is obtained that
\begin{eqnarray}
S\left( y_{A},y_{B}\right) &=&\left( \lambda_{A}-1\right) \left( \lambda_{B}-1\right) \left[
\sum_{n_{1}=0}^{\infty} \sum_{n_{2}=0}^{\infty} 
f_{n_{1},n_{2}}\frac{y_{A}^{n_{1}-1} y_{B}^{n_{2}}}{\left( n_{1}-\lambda_{A}+1\right) \left( n_{2}-\lambda_{B}+1\right)}
\right. \nonumber\\
&&  - \sum_{n_{1}=0}^{\infty} I_{2}\left( \lambda_{B},n_{1}\right) 
\frac{y_{A}^{ n_{1}-1} y_{B}^{\lambda_{B}-1}}{ n_{1}-\lambda_{A} +1} 
 - \sum_{n_{2}=0}^{\infty} I_{1}\left( \lambda_{A},n_{2}\right) 
\frac{y_{A}^{\lambda_{A}-2}y_{B}^{n_{2}}}{ n_{2}-\lambda_{B} +1} \nonumber\\
&& \left. +  I\left( \lambda_{A}, \lambda_{B}\right) y_{A}^{\lambda_{A}-2}y_{B}^{\lambda_{B}-1} \right],
\label{Sfinal}
\end{eqnarray}
where
\begin{eqnarray}
I_{2}\left( \lambda_{B},n_{1}\right)&=& \sum_{n_{2}=m_{2}} ^{\infty}f_{n_{1},n_{2}} \int_{0}^{\infty} x_{2}^{n_{2}-\lambda_{B}} dx_{2}
\nonumber\\
&=& \int_{0}^{\infty} x_{2}^{-\lambda_{B}}
\left(  \frac{1}{n_{1}!} \left. \frac{\partial^{n_{1}}F}{\partial x_{1}^{n_{1}}} \left( x_{1},x_{2} \right) \right|_{x_{1}=0}
- \sum_{n_{2}=0}^{m_{2}-1} f_{n_{1},n_{2}}x_{2}^{n_{2}} \right) dx_{2}, \nonumber\\
\label{I2}
&&
\end{eqnarray}
\begin{eqnarray}
 I\left( \lambda_{A}, \lambda_{B}\right) &=&
\sum_{n_{1}=m_{1}}^{\infty} \sum_{n_{2}=m_{2}}^{\infty} f_{n_{1},n_{2}} 
\int_{y_{A}}^{\infty} x_{1}^{n_{1}-\lambda_{A}} dx_{1} \int_{y_{B}}^{\infty} x_{2}^{n_{2}-\lambda_{B}} dx_{2} \nonumber \\
&=& \int_{0}^{\infty}\int_{0}^{\infty} x_{1}^{-\lambda_{A}}x_{2}^{-\lambda_{B}} \left[ F\left( x_{1},x_{2}\right) -
\sum_{n_{2}=0}^{m_{2}-1} \frac{x_{2}^{n_{2}}}{n_{2}!}  \left. \frac{\partial^{n_{2}}F}{\partial x_{2}^{n_{2}}} \left( x_{1},x_{2} \right) \right|_{x_{2}=0}
\right. \nonumber \\
&& \left.
- \sum_{n_{1}=0}^{m_{1}-1}  
 \frac{x_{1}^{n_{1}}}{n_{1}!} \left. \frac{\partial^{n_{1}}F}{\partial x_{1}^{n_{1}}} \left( x_{1},x_{2} \right) \right|_{x_{1}=0}
+ \sum_{n_{1}=0}^{m_{1}-1} \sum_{n_{2}=0}^{m_{2}-1} f_{n_{1},n_{2}} x_{1}^{n_{1}}x_{2}^{n_{2}} \right] dx_{1}dx_{2}
\nonumber\\
\label{I}
\end{eqnarray}
and the integrals in Eq.\ (\ref{I2}) and Eq.\ (\ref{I}) converge.

In particular, it can be seen that for $F\left( x_{1},x_{2}\right) = x_{1}\tanh \left( x_{1}+x_{2}\right)$ there is
$f_{0,0}=f_{1,0}=0$ and $f_{0,n_{2}}=0$ for $n_{2}=0,1,2\ldots$,
thus also $I_{2}\left( \lambda_{B},0\right)=0$ from Eq.\ (\ref{I2}) and terms corresponding to $n_{1}=0$ (containing $y_{A}^{-1}$)
in the sums in Eq.\ (\ref{Sfinal}) disappear.

\end{document}